\documentclass[prx,amsmath,amssymb, twocolumn,longbibliography]{revtex4-1}
\usepackage{graphicx}
\usepackage{dcolumn}
\usepackage{bm}
\usepackage{placeins}
\usepackage{rotating}
\usepackage{multirow}
\usepackage{longtable}
\usepackage[table]{xcolor}
\usepackage[english]{babel}
\usepackage{amssymb} 
\usepackage{amsmath} 
\usepackage[utf8]{inputenc} 

\newcommand{\be}{\begin{equation}}
\newcommand{\ee}{\end{equation}}
\newcommand{\bea}{\begin{eqnarray}}
\newcommand{\eea}{\end{eqnarray}}
\newcommand{\nn} {\nonumber}

\def\a{\alpha}
\def\b{\beta}

\def\d{\delta}
\def\D{\Delta}

\def\ve{\varepsilon}

\def\l{\lambda}
\def\L{\Lambda}

\def\S{\Sigma}

\def\vf{\varphi}

\def\w{\omega}

\def\bra{\langle}
\def\ket{\rangle}

\def\xc{{\rm xc}}
\def\x{{\rm x}}

\begin{document}
\widetext 

\title{Random phase approximation with exchange for an accurate description of \\crystalline polymorphism}
\author{Maria Hellgren}
\affiliation{Sorbonne Universit\'e, MNHN, UMR CNRS 7590, IMPMC, 4 place Jussieu, 75005 Paris, France}
\author{Lucas Baguet}
\affiliation{Sorbonne Universit\'e, MNHN, UMR CNRS 7590, IMPMC, 4 place Jussieu, 75005 Paris, France}
\date{\today}
\begin{abstract}
We determine the correlation energy of BN, SiO$_2$ and ice polymorphs employing a recently developed RPAx (random phase approximation with exchange) approach. The RPAx provides larger and more accurate polarizabilities as compared to the RPA, and captures effects of anisotropy. In turn, the correlation energy, defined as an integral over the density-density response function, gives improved binding energies without the need for error cancellation. Here, we demonstrate that these features are crucial for predicting the relative energies between low- and high-pressure polymorphs of different coordination number as, e.g., between $\a$-quartz and stishovite in SiO$_2$, and layered and cubic BN. Furthermore, a reliable (H$_2$O)$_2$ potential energy surface is obtained, necessary for describing the various phases of ice. The RPAx gives results comparable to other high-level methods such as coupled cluster and quantum Monte Carlo, also in cases where the RPA breaks down. Although higher computational cost than RPA we observe a faster convergence with respect to the number of eigenvalues in the response function.
\end{abstract}
\maketitle
\section{Introduction}
The development of an accurate, yet computationally efficient, electronic structure approach able to 
treat electron correlation in solids remains an important task in condensed matter physics. Density 
functional theory (DFT) provides a rigorous and computationally appealing framework \cite{hk,ks} that has been widely successful, but available 
approximations still suffer a number of drawbacks. The van der Waals (vdW) interactions, ubiquitous in systems ranging from layered materials to superconducting hydrides, are challenging to include \cite{bjorkman}, and most exchange-correlation (xc) 
functionals do not achieve the accuracy needed to satisfactorily describe
the energetics of phase transformations and polymorphism, even in common systems like water \cite{waterreview} and silica \cite{haysio}. 

The random phase approximation (RPA) represents the highest level of 
sophistication currently applied in materials science \cite{gironcolirpa,rpaprl,ren2012random}. Being based on the exact 
adiabatic-connection fluctuation-dissipation (ACFD) formula \cite{acfd,helgaker2011}, first-order Hartree-Fock exchange is exactly 
incorporated and vdW forces are seamlessly built in, providing an overall improved accuracy. 
However, the performance of the RPA strongly relies on error cancellation, leading to unpredictable errors 
when dealing with crystals of low symmetry \cite{perdewsio,ruzrpa}. Futhermore, vdW forces are underestimated by an average of 20\% \cite{rparangewater,rocca,hcdg18}. The origin of these errors can, however, easily be traced to the approximate Hartree response 
function used for constructing the correlation energy. 

The combination of the ACFD formula and time-dependent (TD) DFT has opened a path for systematic improvements of the RPA \cite{lein,tddft}. 
Within TDDFT, xc effects are incorporated via the xc kernel that can be defined as the second density variation of the xc action functional \cite{fxc,keldyshaction,vbdvls05}. Starting from the RPA, i.e., the Hartree kernel, further refinements then naturally proceed via the local density approximation (LDA) and 
generalized gradient approximations (GGAs). However, first results found that these approximations can worsen the performance with respect 
to the RPA \cite{furche}. A renormalized LDA kernel that introduces spacial nonlocality was, therefore, formulated in Ref. \cite{olsen}. This kernel, 
as well as a renormalized PBE (Perdew-Burke-Ernzerhof) kernel, have shown to improve the RPA correlation energy, leading to a better 
performance in several cases \cite{olsen2,ruzrpa,olsen3}.

Within an approach that combines many-body perturbation theory (MBPT) and TDDFT the first step beyond RPA is to include the full Fock exchange term in the 
density response function via the nonlocal and frequency dependent exact-exchange (EXX) kernel. This generates the random phase approximation with exchange (RPAx), 
that exactly includes the second order exchange diagram as well as higher order exchange effects, in addition to the RPA ring series of diagrams \cite{fetter,PhysRevA.57.3433,hvb08,hvb10,gorling}. 
Various partial resummations of the RPAx correlation terms are thus equal to other advanced expressions for the correlation energy 
defined within MBPT such as, e.g., SOSEX (second order screened exchange) \cite{sosex1,sosexkresse,jansensosex,sosex2,chdg14,hcdg18}. First tests on atoms and molecules have shown that not only accurate correlation energies are obtained with RPAx \cite{chdg16,hcdg18}, 
but also xc potentials, i.e., electronic densities, and polarisabilities \cite{hvb10,scrpaxgorling}. 
Furthermore, studies on the homogeneous electron gas, the simplest model of a metal, 
have shown promising results \cite{chdg14}. 
 
In this work, we extend the scope of applications to solids and show that it is possible to reach results of similar quality 
as for molecules. We calculate the relative energies between various crystalline phases within the BN, SiO$_2$ and ice polymorphs. 
In particular, we investigate cases where the RPA has shown insufficient, as, e.g., for the $\a$-quartz-stishovite energy difference 
in SiO$_2$, for which the change in coordination number necessitate accurate correlation energies \cite{perdewsio,haysio,ruzrpa}. 
A similar problem occurs in cubic and layered BN, inverting their stability order \cite{gouldbn,ccsdtbn}. 
In both SiO$_2$ and BN the vdW forces play an important role \cite{marinibn,bjorkman,haysio}. Here we investigate the effect of the 
EXX kernel on the binding energy of layered BN as well as the energy difference between $\a$-quartz and cristobalite. A study 
of purely vdW bonded solids, such as Ar and Kr, confirms that the vdW bond is well described with RPAx, 
yielding an accuracy similar to that found for molecular dimers. 
Finally, we perform a detailed analysis of the water dimer potential energy surface and ice polymorphism. The delicate 
balance between Pauli repulsion, static and dynamic correlation has made water a long-standing problem for electronic 
structure methods. 

Our results on this broad and challenging set of systems show promise for future applications. The RPAx stays within the simple 
computational framework of the RPA, but has an accuracy comparable with more sophisticated methods such as quantum Monte Carlo 
(QMC) and coupled cluster (CC). 
\begin{figure}[t]
\includegraphics[scale=0.23]{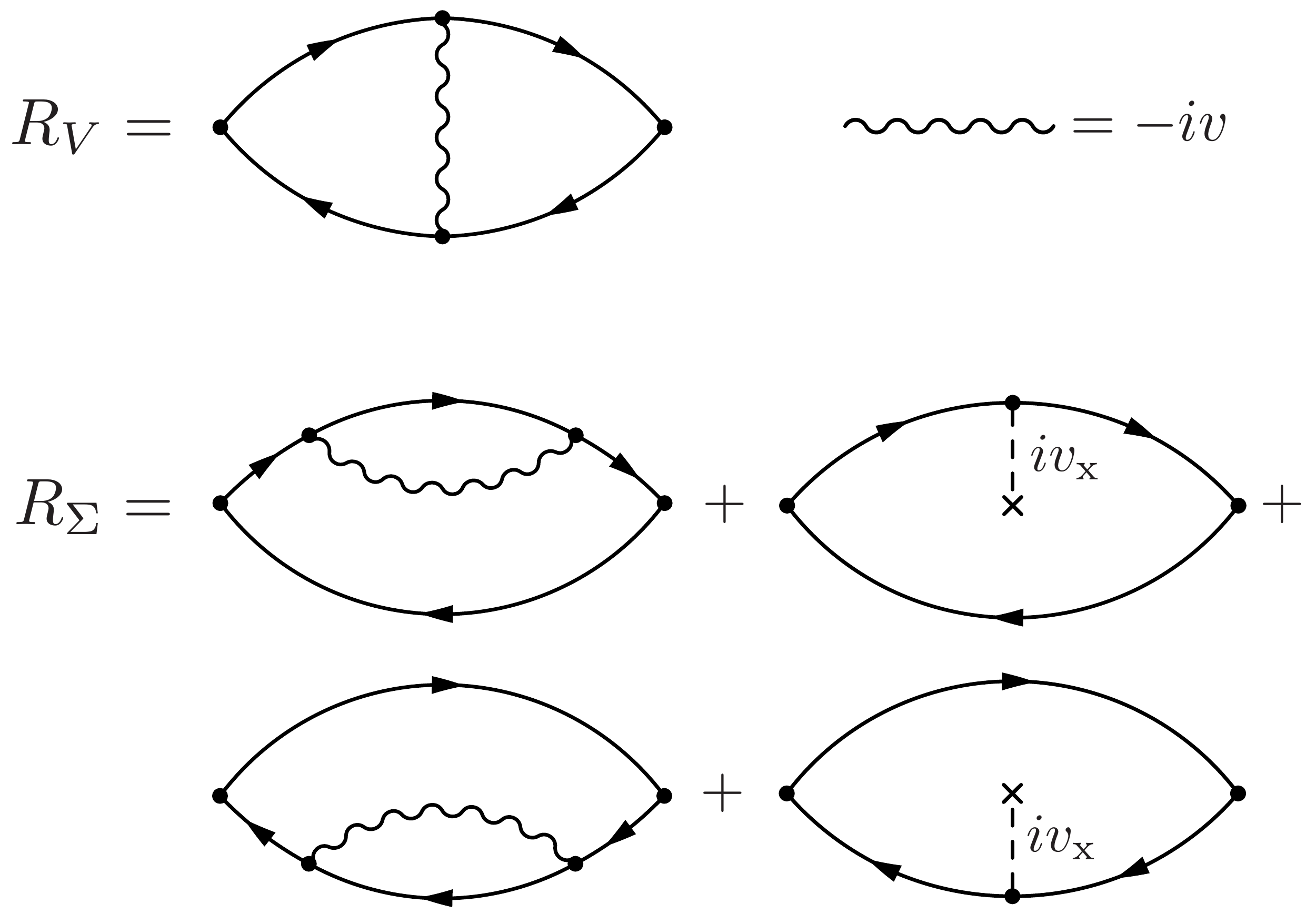}
\caption{Diagrammatic representation of the two terms in Eq. (\ref{fx}). Full lines correspond to KS Green functions.}
\label{diagrams}
\end{figure}

\section{RPA with exchange}
In this section we summarize the equations of the RPAx approximation for periodic solids. 
These have been introduced previously \cite{hvb08,hvb10} but are here adapted to the specific implementation employed in this work \cite{chdg14,chdg16,hcdg18}. 

The RPAx approximation for the correlation energy is based on the ACFD (adiabatic connection fluctuation-dissipation) formula, 
in which the exact correlation 
energy is expressed in terms of the linear density response function, $\chi_{\lambda}$, multiplied by the 
Coulomb interaction, $v$, and integrated over the frequency and the interaction strength $\l$
\be
E_{\rm c}= -\int_0^1 d\lambda\int_0^\infty \frac{d\w}{2\pi}{\rm Tr}\{v[\chi_{\lambda}(i\omega)-\chi_s(i\omega)]\}.
\label{ecorr}
\ee 
The underlying Hamiltonian is given by 
\be
H=T+ V^\l+\lambda W,
\ee
where $T$ is the kinetic energy operator, $W$ is the electron-electron interaction operator, 
and $V^\l$ is a one-body potential operator composed of the external potential and a 
fictitious potential that fix the density at full interaction 
strength for every $\l$. With this definition, the system at $\l=0$ already produces the exact density, $n$, via 
the exact Kohn-Sham (KS) xc potential, $v_\xc$. The KS independent-particle response function, 
subtracted in Eq. (\ref{ecorr}), is denoted $\chi_s$.

Within TDDFT, $\chi_\l$ becomes a functional of the ground-state density, $n_0$, 
via the xc kernel, $f^{\l}_\xc=\d v^{\l}_\xc/\d n|_{n_0}$. 
This can be expressed in a Dyson-like equation  
\be
\chi_{\lambda}(\omega)=\chi_{s}(\omega)+\chi_s(\omega)[\l v+f^{\l}_\xc(\w)]\chi_{\lambda}(\omega),
\label{dyson}
\ee
similar to the RPA (or Hartree) equation ($f_\xc=0$). 

A systematic approach for generating xc kernels based on many-body perturbation theory was 
introduced in Ref. \onlinecite{vbdvls05}. Within this approach the simplest approximation 
corresponds to including the full Fock exchange term via a time-dependent optimized effective 
potential scheme. This results in the EXX kernel, $f_\x$, defined by the 
following equation 
\be
\chi_s(\w)f_\x(\w)\chi_s(\w)=R_V(\w)+R_{\S}(\w).
\label{fx}
\ee
The right hand side can be interpreted diagrammatically in terms of KS Green functions, being the sum of 
the first order vertex diagram, $R_V(\w)$, and the first order self-energy correction diagrams
$R_{\S}(\w)$ (see Fig.~\ref{diagrams}). Although $f_\x$ is first order in the Coulomb interaction, the correlation 
energy, via Eq.~(\ref{ecorr}), contains all orders in $v$. This approximation to the 
correlation energy is what we call RPAx and is the same as the one used in Ref. \cite{gorling}. Other variants based on a dielectric formulation \cite {dielect}, or Hartree-Fock orbitals and range-separation 
have also been proposed \cite{toulouserpax}. We emphasize that the RPAx, as defined here, is based 
on TDDFT and KS orbitals, thus falling in the category of orbital functionals in DFT (i.e. functionals depending on the 
(un)occupied KS orbitals and eigenvalues). The total energy functional can then be minimized via a local KS potential in the usual way \cite{hvb10}.

The equation for the EXX kernel, Eq.~(\ref{fx}), is usually expressed in terms of 
KS orbitals (occupied and unoccupied), $\vf_n$, and KS eigenvalues, $\ve_n$. In Ref. \onlinecite{chdg14} it was shown that the same expression can be reformulated 
in terms of occupied KS orbitals and their responses. For periodic systems this expression reads
\begin{widetext}
\bea
R_{V}^{\a\b}({\bf q},i\w)&=&-\sum_{{\bf k}n{\bf k'}n'}^{\rm occ}\bra \D^\a \vf_{n{\bf k+q}}^+ \vf_{n'{\bf k'}}|v| \D^{\b} \vf_{n'{\bf k'+q}}^-\vf_{n{\bf k}} \ket -\sum_{{\bf k}n{\bf k'}n'}^{\rm occ}\bra \D^\a \vf_{n{\bf k+q}}^- \vf_{n'{\bf k'}}|v| \D^{\b} \vf_{n'{\bf k'+q}}^+\vf_{n{\bf k}} \ket \nn\\
&&-\sum_{{\bf k}n{\bf k'}n'}^{\rm occ}\bra \D^{\a} \vf_{n{\bf k+q}}^-  \vf_{n(-{\bf k})} |v|\vf_{n'(-{\bf k'})} \D^\b \vf_{n'{\bf k'+q}}^-\ket-\sum_{{\bf k}n{\bf k'}n'}^{\rm occ}\bra \D^{\a} \vf_{n{\bf k+q}}^+ \vf_{n(-{\bf k})} |v|\vf_{n'(-{\bf k'})}  \D^\b \vf_{n'{\bf k'+q}}^+\ket
\eea
\bea
R_{\S}^{\a\b}({\bf q},i\w)&=&
-\sum_{{\bf k}nn'}^{\rm occ}\bra \D^\S\vf_{n{\bf k+q}}|\D^{\b} \vf_{n'{\bf k+q}}^++\D^{\b} \vf_{n'{\bf k+q}}^-\ket \bra  \vf_{n'{\bf k}}|V_{\bf -q}^{\a}| \vf_{n{\bf k+q}} \ket+\sum_{{\bf k}n}^{\rm occ}\bra   \D^\S \vf_{n{\bf k}} |V^{\a }_{\bf -q}|\D^{\b} \vf_{n{\bf k+q}}^++ \D^{\b} \vf_{n{\bf k+q}}^- \ket\nn\\
&&-\sum_{{\bf k}nn'}^{\rm occ}\bra \D^{\a} \vf_{n{\bf k+q}}^++\D^{\a} \vf_{n{\bf k+q}}^-| \D^\S\vf_{n'{\bf k+q}}\ket \bra  \vf_{n'{\bf k+q}}|V_{\bf q}^{\b }| \vf_{n{\bf k}} \ket+\sum_{{\bf k}n}^{\rm occ}\bra \D^{\a} \vf_{n{\bf k+q}}^++ \D^{\a} \vf_{n{\bf k+q}}^- |V^{\b}_{\bf q}|   \D^\S \vf_{n{\bf k}}\ket\nn\\
&&-\sum_{{\bf k}nn'}^{\rm occ}\left[\bra \D^{\a} \vf^{-}_{n{\bf k+q}}|\D^{\b} \vf^+_{n'{\bf k+q}} \ket+\bra \D^{\a} \vf^{+}_{n{\bf k+q}}|\D^{\b} \vf^-_{n'{\bf k+q}} \ket\right] \bra  \vf_{n'{\bf k}}|\S_{\bf k}-v_\x| \vf_{n{\bf k}} \ket\nn\\ 
&&+\sum_{{\bf k}n}^{\rm occ}\bra  \D^{\a} \vf_{n{\bf k+q}}^-|\S_{\bf k+q}-v_\x|\D^{\b} \vf_{n{\bf k+q}}^+\ket +\sum_{{\bf k}n}^{\rm occ}\bra  \D^{\a} \vf_{n{\bf k+q}}^+|\S_{\bf k+q}-v_\x|\D^{\b} \vf_{n{\bf k+q}}^-\ket 
\eea
\end{widetext}
where
\be
 |\D^\a \vf^\pm_{{\bf k+q},n}\ket= \sum_m^{\rm unocc}|\vf_{{\bf k+q},m}\ket\frac{\bra\vf_{{\bf k+q},m}|V_{\bf q}^\a|\vf_{{\bf k},n}\ket}{\pm iw+\ve_{{\bf k},n}-\ve_{{\bf k+q},m}}
\ee
\be
 |\D^{\S} \vf_{{\bf k},n}\ket= \sum_m^{\rm unocc}|\vf_{{\bf k},m}\ket\frac{\bra\vf_{{\bf k},m}|\S_{{\bf k}}-v_{\x}|\vf_{{\bf k},n}\ket}{\ve_{{\bf k},n}-\ve_{{\bf k},m}}
\ee
with $\S_{{\bf k}}$($v_{\x}$) the nonlocal(local) Fock exchange potential and $V^\a_{\bf q}$ a local potential. In this form 
one can apply the linear response techniques of density functional perturbation theory to determine Eq.~(\ref{fx}) 
and, hence, avoid the generation of unoccupied KS states. 

By solving the generalized eigenvalue problem
\be
 R|u_\b\ket=\nu_\b\chi_s|u_\b\ket
\label{eig}
\ee
where $R=R_V+R_{\S}+\chi_s v\chi_s$, the correlation energy can be computed from
\bea
E_{\rm c}&=&-\frac{1}{N_{\bf q}}\sum_{\bf q}\sum^{N_\nu}_\b\int_0^\infty \frac{d\w}{2\pi}\frac{\bra u_\b|\chi_s v\chi_s|u_\b \ket}{\nu_\b({\bf q},i\w)}\nn\\
&&\,\,\,\,\,\,\,\,\,\,\,\,\,\,\,\,\,\,\,\,\,\,\times\{\nu_\b({\bf q},i\w)+\ln[1-\nu_\b({\bf q},i\w)]\}.
\label{echi}
\eea
where $N_\nu$ is the total number of eigenvalues and $N_{\bf q}$ is the total number of $\bf q$-points in the Brillouin zone. The latter can be reduced exploiting crystal symmetries. The $\l$-integration is done analytically, while the frequency integral has to be carried 
out numerically. We immediately see that this expression is reduced to RPA by setting $R_V=0$ and $R_{\S}=0$ in Eq. (\ref{eig}). 

\begin{figure*}[t]
\includegraphics[scale=0.27,angle=90]{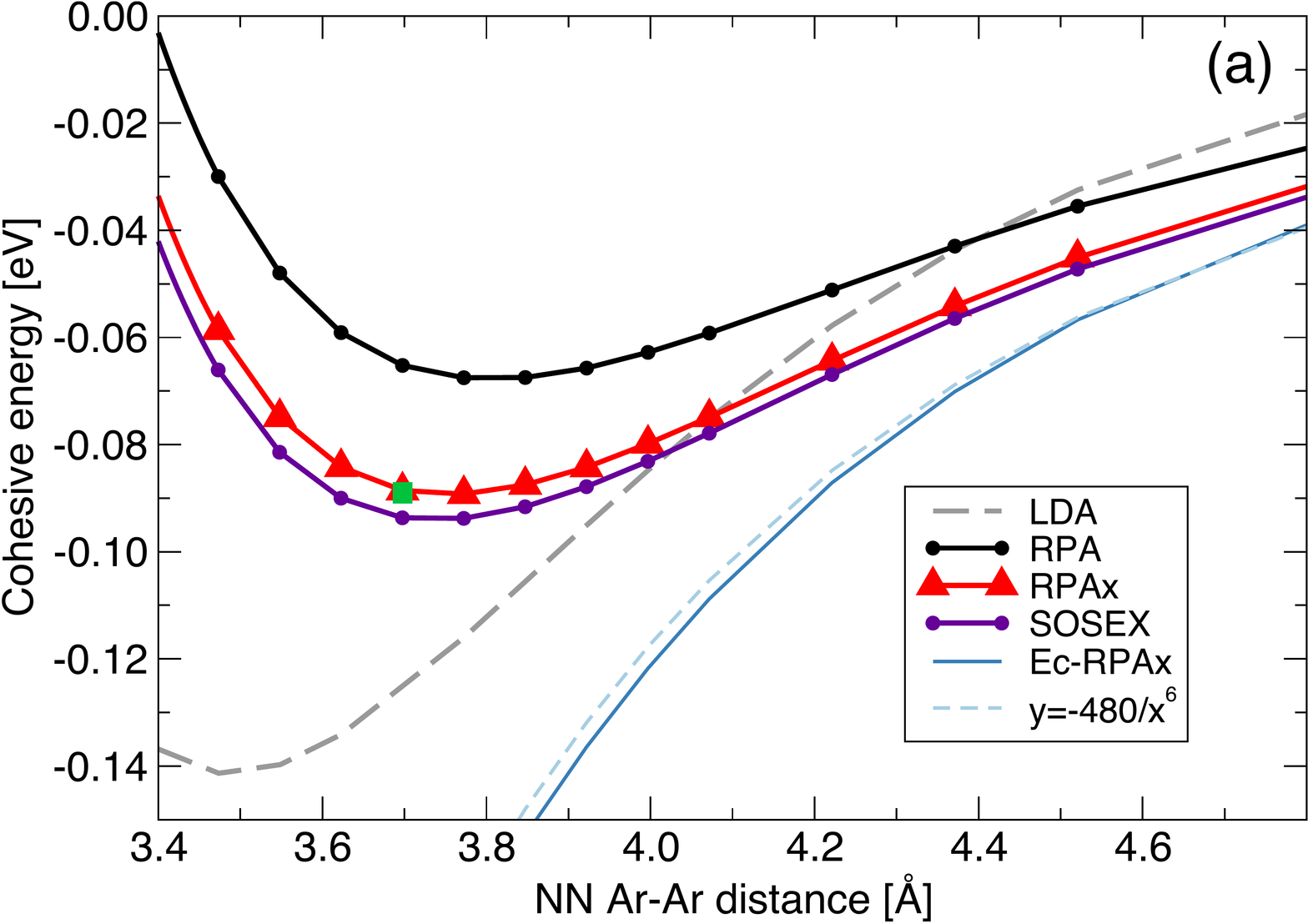}
\hspace{0.8cm}
\includegraphics[scale=0.27,angle=90]{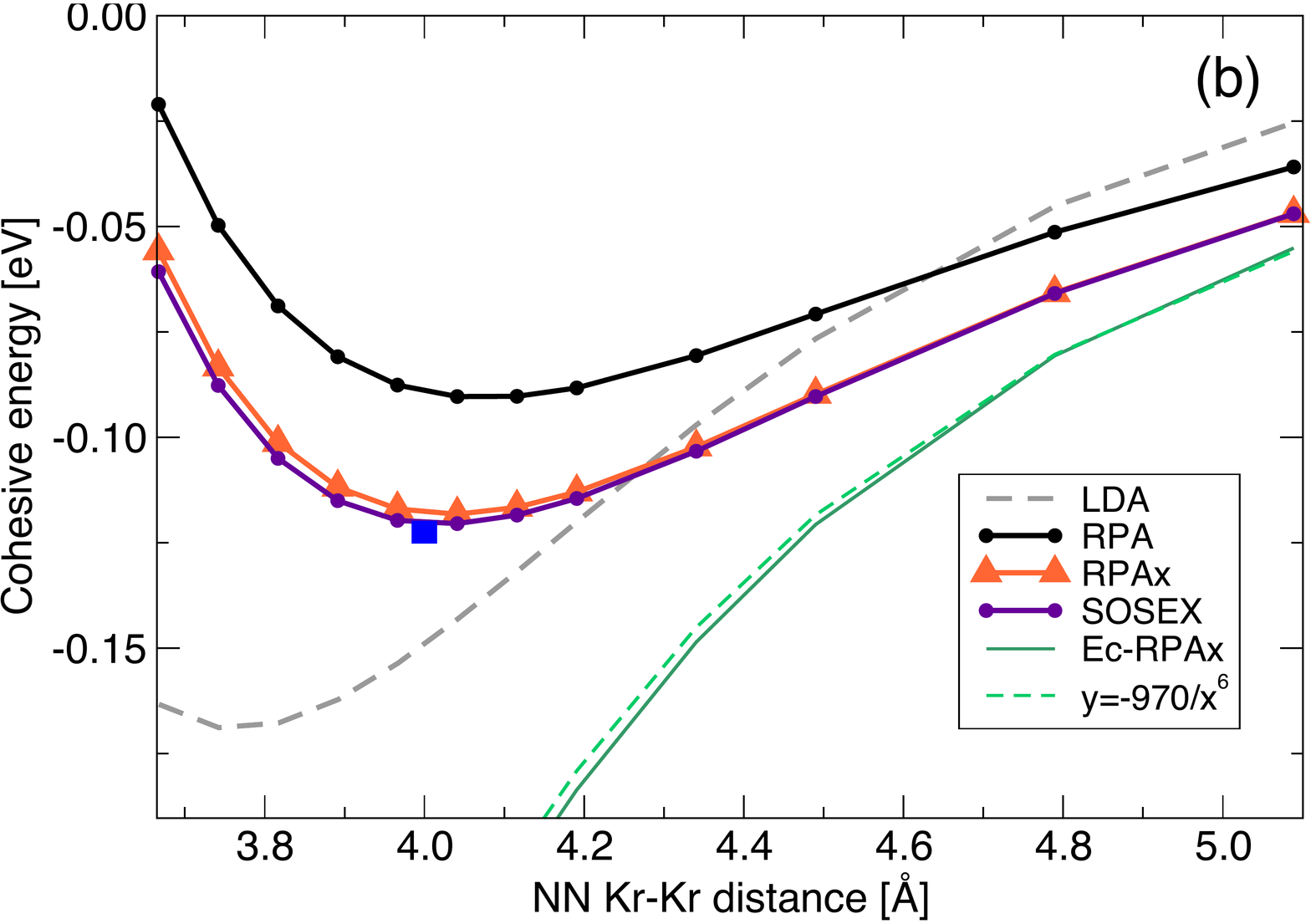}
\caption{Cohesive energy (eV) as a function of nearest neighbour distance (\AA) in solid Ar (a) and Kr (b). The correlation energy (Ec-RPAx) is also shown, together with a fitted $y=k/x^6$ curve. Green and blue squares correspond to CCSD(T) results \cite{stoll99}.}
\label{krar}
\end{figure*}
Equations~(5)-(10) have been implemented within the plane-wave and pseudopotential framework, but have so far only been applied to 
molecular systems \cite{chdg14,chdg16,hcdg18}. Binding energies have been shown to converge 
quickly with respect to the number of eigenvalues ($\approx10$ times the number of electrons ($N_e$)). 
Due to the additional $k$-point sum in $R_V$, RPAx scales as $N_k^3$ with respect to the number of 
k-points, which can be compared to $N_k^2$ with RPA. On the other hand, as we will see later, the RPAx converges, in many cases, 
faster than RPA with respect to $N_\nu$. 

The underlying many-body formulation of the $f_\x$-kernel allows also a variant of SOSEX \cite{sosex1,sosexkresse,jansensosex,sosex2}
to be generated within the present formalism, making use of an alternative resummation of Dyson's equation (Eq. (\ref{dyson})), that only includes a subset of terms \cite{hcdg18}. In the following we will present both RPAx and SOSEX results.

Since a fully self-consistent implementation is still not developed, all calculations are done on top of a PBE ground-state \cite{pbe}, unless otherwise stated. The electron-ion interaction is treated with standard norm-conserving pseudopotentials \cite{pseudo1,PhysRevB.54.1703,pseudo2}.   
\section{Noble gas solids}
The ACFD formula for the correlation energy provides a natural starting-point for including the long-range vdW force, i.e., the attractive force between two charge neutral, closed-shell atoms $A$ and $B$ due to fluctuating dipoles present in a correlated wave function. 
At large separation $R$, the interaction energy can, to leading order, be written as \cite{zk76}
\be
E^{AB}=-\frac{1}{R^6}\frac{3}{\pi}\int_0^\infty \!d\w \,\a_{\rm iso}^A(i\w) \a_{\rm iso}^B(i\w)\equiv-\frac{C_6}{R^6},
\label{e2}
\ee 
where
\be
\a_{\rm iso}^A(i\w)=-\frac{1}{3}\sum_i\int \!d^3{\bf r} d^3{\bf r'} x_i \,\chi^A({\bf r},{\bf r'},i\w)x'_i 
\label{isoa}
\ee 
is the isotropic dynamical polarizability of atom $A$, and $C_6$ is the vdW coefficient. In Ref.~\onlinecite{Dobson}, Dobson showed that the RPA correlation energy 
exactly reproduces Eq. (\ref{e2}), but with the exact polarizability replaced by the RPA polarizability. 
However, as shown in several previous works, the RPA $C_6$-coefficients are underestimated (up to 40\% in highly polarizable atoms) \cite{hvb10,toulouse2,gould2012,PhysRevB.88.115131,gould2013,sosex2}. 
For noble gas atoms the error lies at about 17\%,  which induces an error of approximately 30\% in the binding 
energy of dimers \cite{toulouse2,chdg16}. Thus, while the RPA captures the vdW forces in 
a qualitatively correct way, the quantitative errors are large. 

Including the xc kernel, it is not straightforward to derive an expression similar to Eq. \ref{e2}. In fact, in Ref. \cite{gould_sc} it was 
shown that semilocal kernels produce an additional $1/R^6$-term. It remains to be proven analytically whether the 
nonlocal RPAx reduces to Eq. \ref{e2}.

In Fig.~\ref{krar} we have plotted the cohesive energy of solid Ar and Kr as a function of nearest 
neighbour (NN) distance. In RPA and RPAx a reduced $3\times3\times3$ shifted $k$-point grid is used for the 
correlation energy giving results within 4 meV of the fully converged values (see Appendix C).  

In the RPA, the cohesive energy is underestimated by 25\%. Our results thus differ somewhat from those obtained in 
Ref. \cite{hk08}, but are more consistent with the results found for dimers (see Refs. \cite{toulouse2,chdg16}). 
As expected, based on the accurate $C_6$-coefficients and dimer binding energies, we find RPAx to be in good agreement with 
both CCSD(T) \cite{stoll99} and experimental results \cite{aziz}. SOSEX gives only slightly larger binding energies. 

In Fig.~\ref{krar} we also display the correlation energy (E$_{\rm c}$-RPAx) as a function of NN distance, exhibiting a 
perfect $1/R^6$ behaviour. Given that Ar and Kr have 12 NN the fitted coefficients 67*12  
and 135*12 (atomic units) compare well with the dimer $C_6$-coefficients of 64 and 130, respectively. Similar accuracy for 
the $C_6$-coefficients are found with other variants of RPAx \cite{gould2013,sosex2,gouldbucko,hesselman_rev} and 
the renormalized LDA kernel \cite{PhysRevB.88.115131}. We note that the SOSEX defined in Ref. \cite{sosex2} is different 
from the one defined here since it does not contain self-energy corrections (here included via $R_\S$ in Eq. (\ref{fx}) \cite{hcdg18}).  
The self-energy corrections are similar to the 'single excitations' in Ref. \cite{sosex2}.
\section{Boron nitride and silica}
\begin{figure}[t]
\includegraphics[scale=0.45]{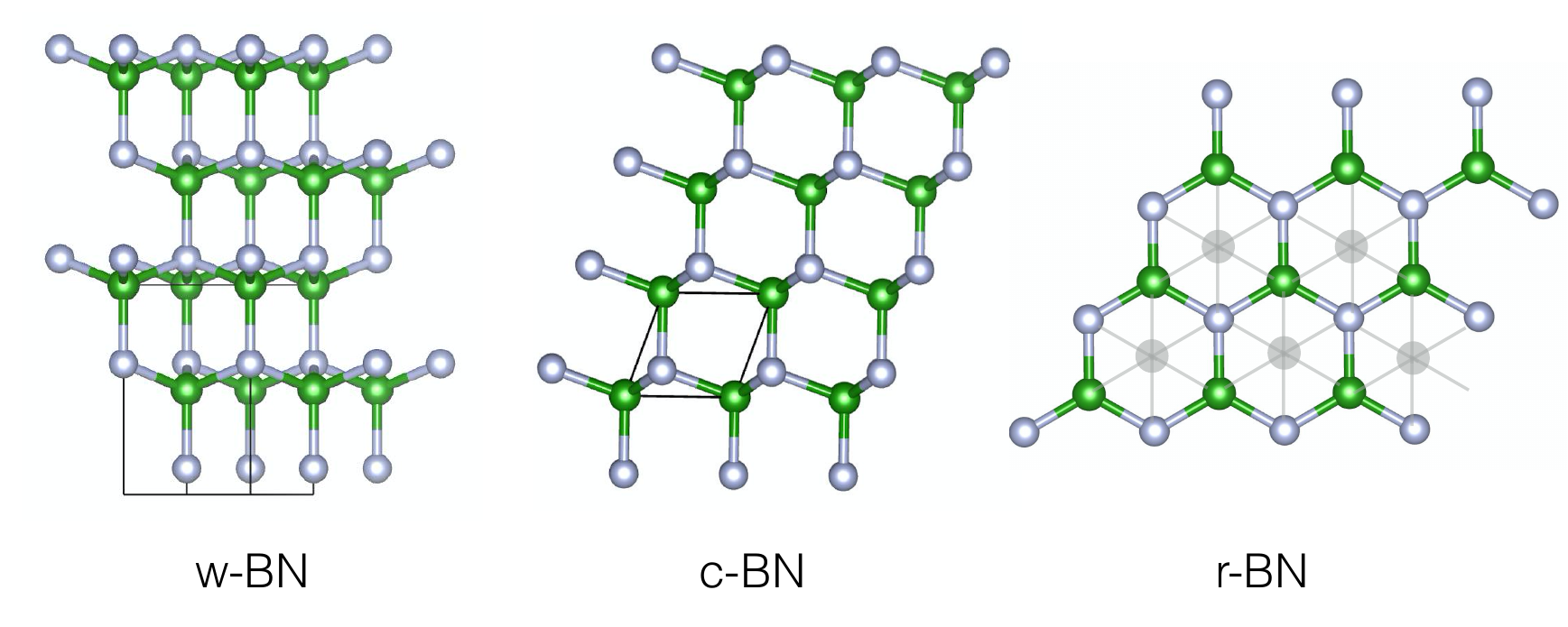}
\includegraphics[scale=0.45]{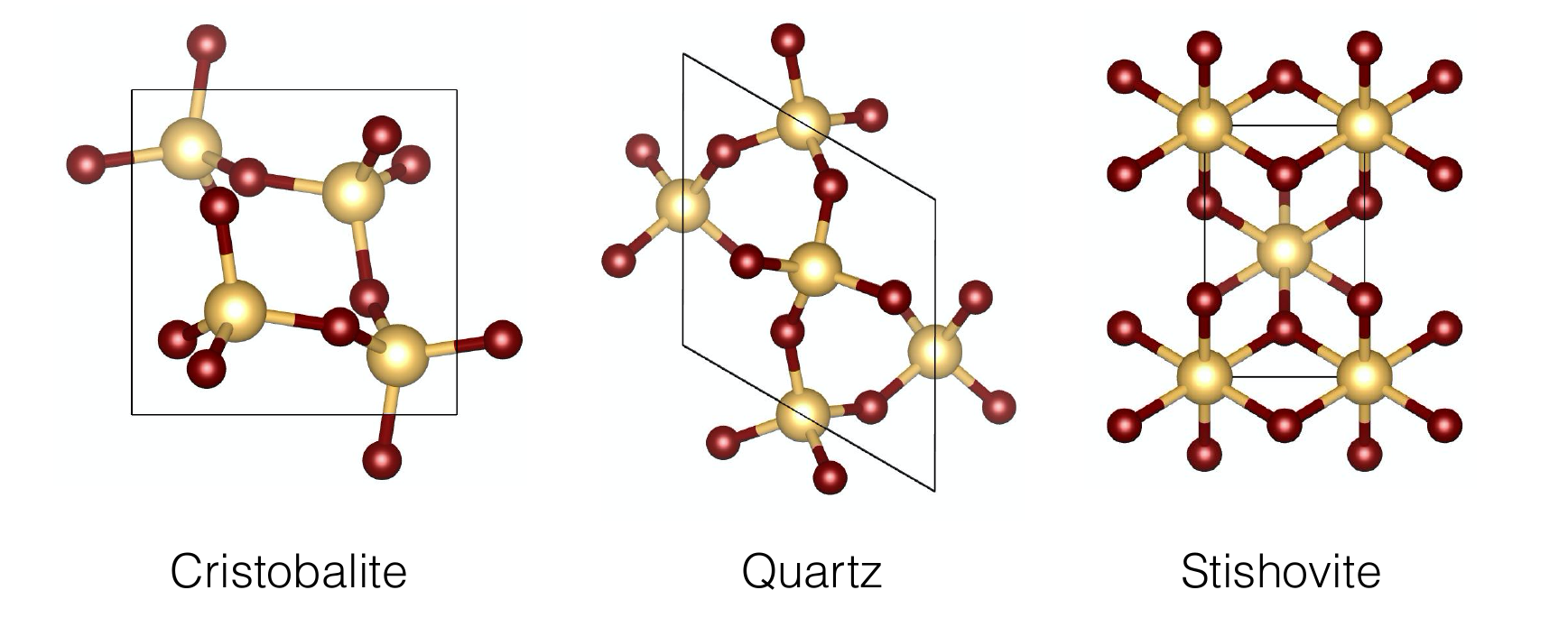}
\caption{Upper row: wurtzite (w-BN), cubic (c-BN) and layered rhombohedral (r-BN) boron nitride. Lower row: cristobalite and $\a$-quartz 
with fourfold coordinated silicon, and stishovite with sixfold coordinated silicon. Silicon in gold and oxygen in red.}
\label{sipoly}
\end{figure}

Having established that vdW bonded solids are accurately reproduced with RPAx we now turn to systems were 
both vdW forces and covalent/ionic bonding are important. The polymorphs of BN 
consist of the high-pressure cubic (c) and wurtzite (w) structures with fourfold coordinated B/N atoms, 
and several low-pressure layered structures, such as hexagonal (h) and rhombohedral (r) BN, which are threefold coordinated (see Fig. \ref{sipoly}). The density, symmetry and ionic 
character of the B-N bond change from one phase to the other, with the largest difference between 
fourfold and threefold coordinated crystals. In these cases, functionals that largely rely 
on error cancellation become unreliable, producing substantial errors in energy differences that can invert the stability order. Although it is well known that the layered polymorphs need a good description of the vdW forces \cite{marinibn}, this is not the only missing component. 
Inaccuracies in describing the covalent bond will show up if errors fail to cancel. 

\begin{figure}[b]
\includegraphics[angle=90,scale=0.27]{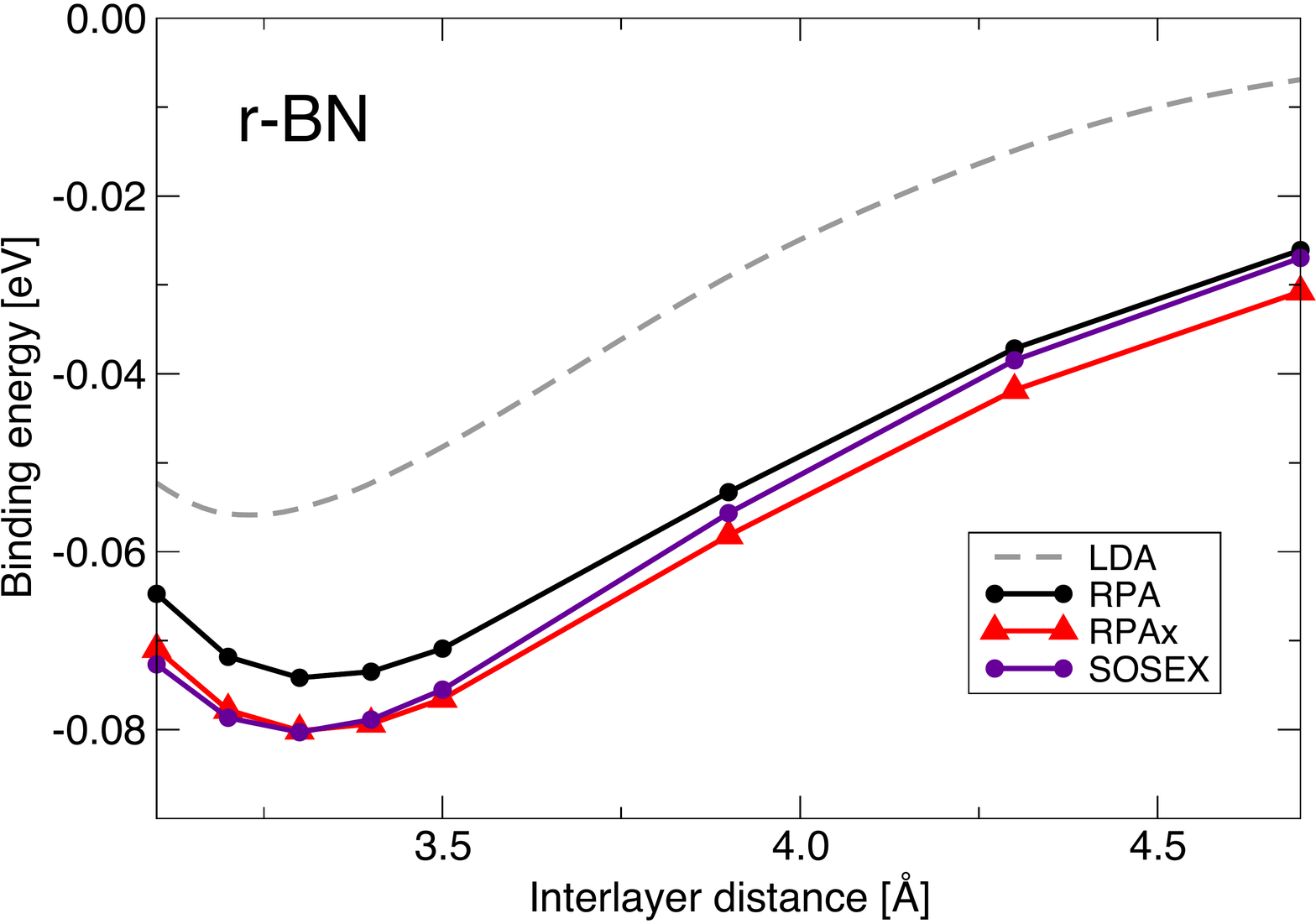}
\caption{Interlayer binding energy (eV) of r-BN as a function of interlayer separation (\AA). }
\label{bndss}
\end{figure}

In Fig. \ref{bndss} we plot the interlayer binding energy of r-BN as a function of interlayer distance, keeping the 
intralayer B-N distance fixed to the experimental value. 
In general, the interlayer bonding is difficult to model with vdW corrected DFT. Binding energies are typically 
largely overestimated, often twice as large as RPA values \cite{bjorkman,vdwbjork}. On the other hand, RPA underestimates 
the vdW forces, but it is not clear to what extent in layered materials. We have verified that we obtain the 
same RPA result as in Ref. \cite{bjorkman}, taking into account that h-BN is 5 meV lower in energy than 
r-BN \cite{gouldbn}. The RPAx increases the binding energy by 7 meV (or 9\%). This correction is smaller than 
the correction in purely vdW bonded solids such as Ar and Kr analyzed above. The RPA thus performs better than 
expected, at least for the BN layered system. We note that, except for some differences in the dissociation tail, 
the SOSEX and RPAx give essentially the same result.

\begin{figure}[t]
\includegraphics[scale=0.4]{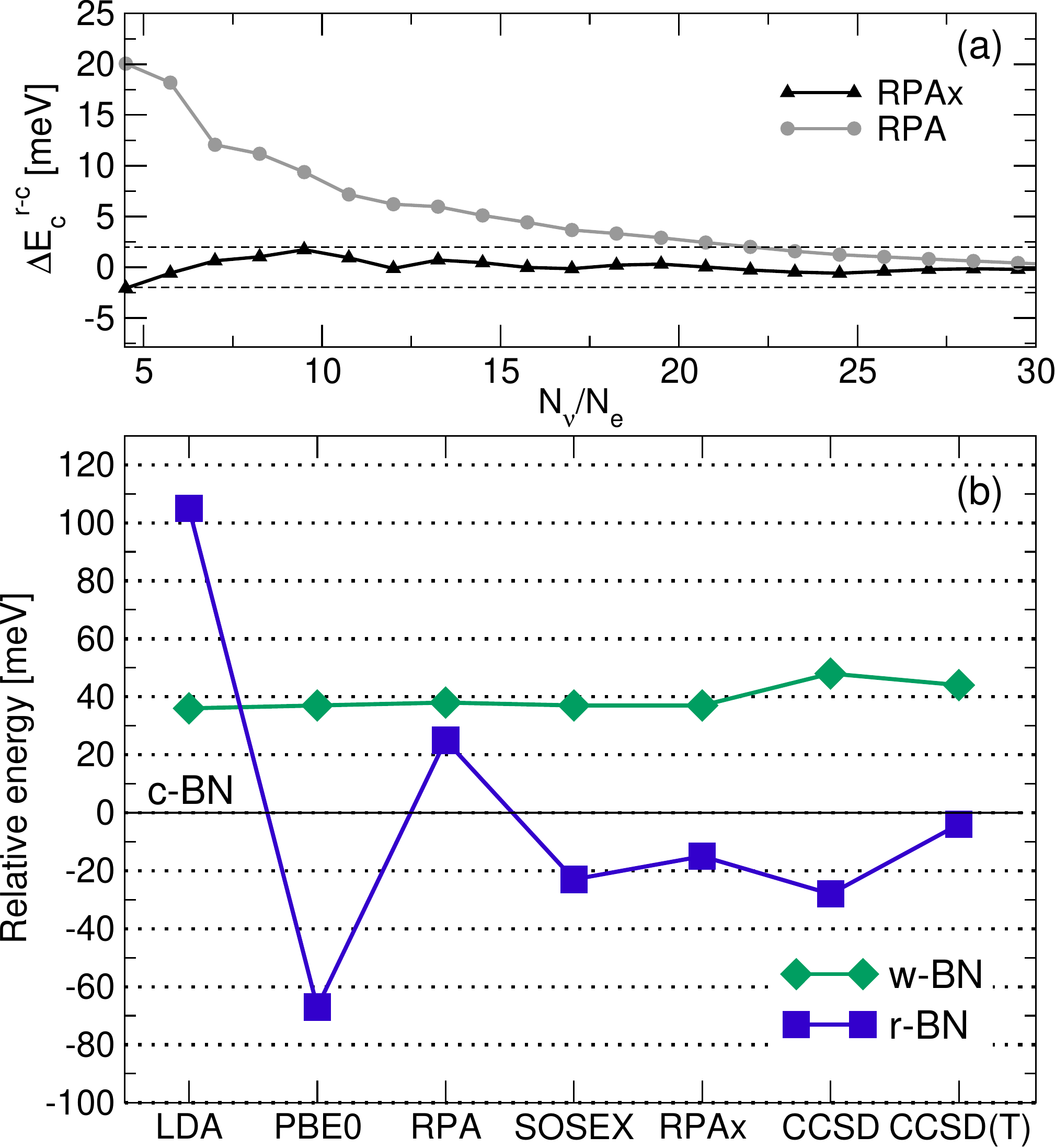}
\caption{(a) Correlation energy difference of r-BN and c-BN as a function of eigenvalues per electron within RPA and RPAx. (b) Relative energies (meV) of w-BN and r-BN with respect to c-BN within LDA, PBE0, RPA, SOSEX and RPAx. CC results are 
from Ref. \onlinecite{ccsdtbn}. c-BN is the reference value at zero.}
\label{bnpoly2}
\end{figure}

In the lower panel of Fig. \ref{bnpoly2} we present results for the energy difference between c-BN and w-BN,  
and between c-BN and r-BN. c-BN is the reference level at zero in the figure. Calculations are done on 
experimental structures \cite{bnexphex}. We compare our RPA and RPAx results to recent CCSD and CCSD(T) results \cite{ccsdtbn}, to LDA and PBE0 (i.e. a hybrid functional with 25\% of Fock exchange). We see a striking difference in the performance on the different polymorphs. 
Both the wurtzite and cubic structures are high-pressure phases with the same B-N coordination and similar volume. This is thus the 
ideal situation for approximations to benefit from error cancellation. Indeed, almost all approximations give the same 
value within a couple of meV. 
The situation is very different when looking at the r-BN and c-BN energy difference. Here, the volume differ by 35\% and 
the B-N bond change character. LDA is seen to hugely overestimate the energy difference (105 meV) and to predict c-BN 
to be lowest in energy. Experimental results are mixed but CC results revert the energy order (see Ref. \cite{ccsdtbn} and experimental references therein \cite{bnexp,FUKUNAGA}). Additionally, PBE0 favours this energy order, predicting r-BN to be lowest 
in energy, although with a too large energy difference (-67 meV). The RPA, that is known to rely on error cancellation, predicts the same energy order as LDA, 
but with a smaller energy difference (25 meV). 
We note that our result is in good agreement with the RPA calculation in Ref. \cite{gouldbn}. The RPAx, known to give 
accurate absolute energies, and not only energy differences, gives a result very close to the CCSD result with an 
energy difference of -15 meV, and thus reverts the stability order. SOSEX gives a slightly larger energy difference (-23 meV). 
In Ref. \cite{ruzrpa} it was shown that the renormalized PBE kernel also provides an improvement upon RPA (-48-(-38) meV). 
We note that the SOSEX defined here is different from the ACSOSEX 
defined in Refs. \cite{acsosex,ruzrpa}. The latter is based on the renormalized PBE kernel whereas the one here is based on many-body 
perturbation theory \cite{hcdg18}.

The difference between RPA and RPAx cannot solely be explained by the difference in how they describe the vdW forces. This suggests that 
the RPAx captures the covalent B-N bond better than RPA.  
Our result provides further confirmation that the correct order places the layered low-density structures at a slightly lower energy than the cubic form, even at zero temperature. The difference in zero-point (vibrational) energy (ZPE), has been estimated to -10 meV \cite{gouldbn}. We also mention that we tried various DFT functionals employing a vdW correction, again with very mixed results. 

In the upper panel of Fig. \ref{bnpoly2} we show the convergence of the correlation energy difference between c-BN and r-BN 
($\D E_{\rm c}^{\rm r-c}$) with respect to $N_\nu/N_{e}$. We note that when comparing the correlation energy of two systems the 
same number of eigenvalues {\em per electron} has to be used \cite{hcdg18}. Since r-BN and c-BN have the same number of 
electrons in the unit cell this implies the same number of eigenvalues. w-BN has twice the number of electrons and, therefore, 
needs twice the number of eigenvalues.  However, while the correlation energy difference between w-BN and c-BN converges 
quickly with respect to the number of eigenvalues ($< 10\times N_e$) the RPA exhibits a slow convergence for the r-BN-c-BN energy 
difference. We used up to 512 eigenvalues in order to produce results within 2 meV (marked by the dashed horizontal lines in Fig. \ref{bnpoly2}), which is more than 60 times the number of 
electrons and hence 6 times more than usual. The problem of not cancelling errors is thus visible also in this technical aspect
of our approach. The absence of this problem is remarkably evident in the convergence of the RPAx correlation energy difference. 
In RPAx we needed $< 64$ eigenvalues, which is similar to the number used for other systems. Further convergence studies 
can be found in Appendix B. 
\begin{figure}[t]
\includegraphics[scale=0.4]{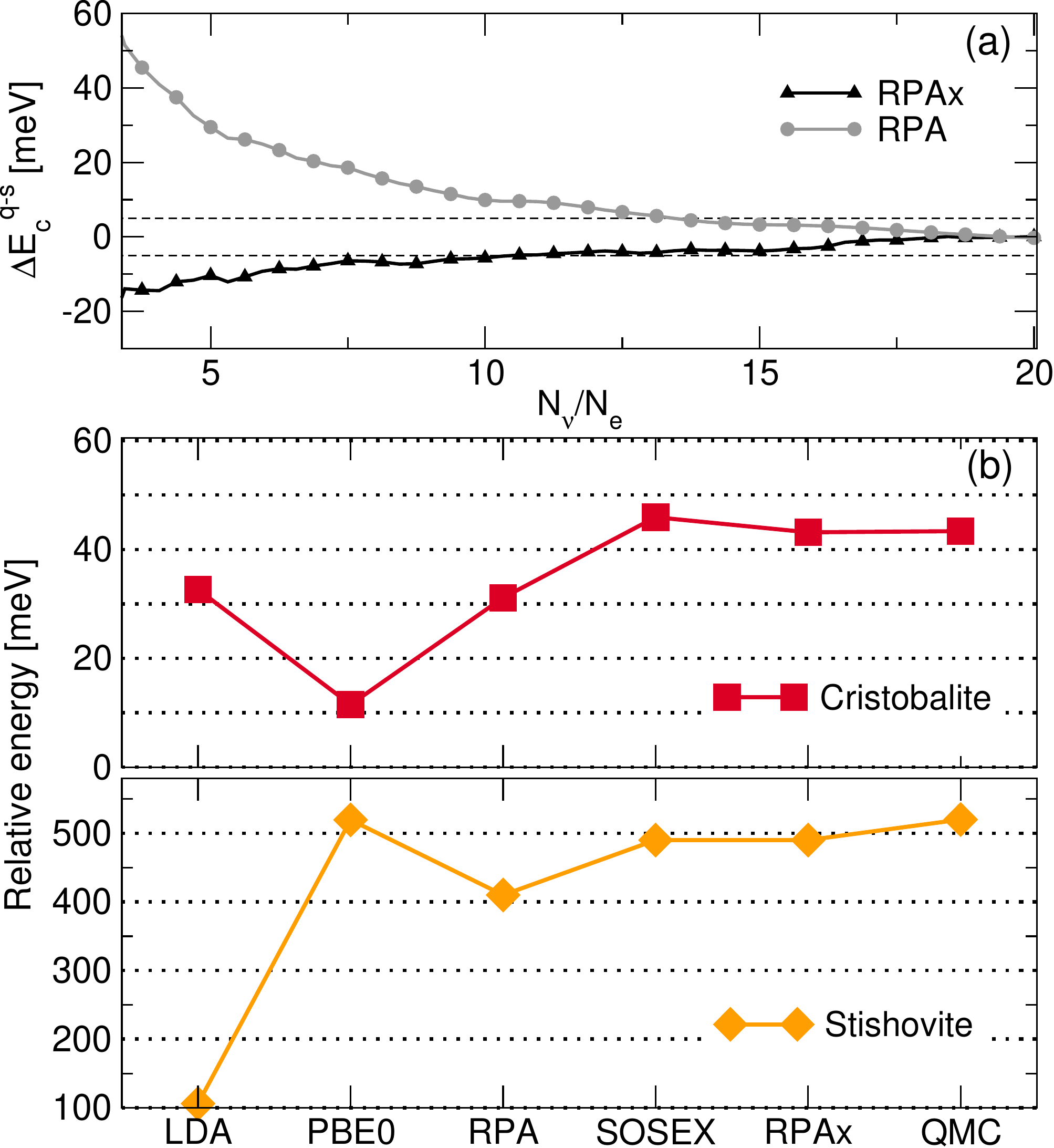}
\caption{(a) Correlation energy difference of stishovite and $\a$-quartz as a function of eigenvalues per electron within RPA and RPAx. 
(b) Relative energies (meV) between cristobalite and $\a$-quartz (red) and between stishovite and $\a$-quartz (yellow) within LDA, PBE0, 
RPA, SOSEX and RPAx. QMC results are from Refs. \onlinecite{qmcsio,haysio}.}
\label{sipoly1}
\end{figure}

A similar effect, as studied above in BN, has previously been observed in silica (SiO$_2$) polymorphs \cite{perdewsio,haysio,ruzrpa}. 
Silica, similar to BN, has polar covalent bonds and polymorphs of different coordination number. In $\a$-quartz (space group P3$_1$21), the most 
stable crystal at ambient pressure, silicon is fourfold coordinated by oxygen, forming a tetrahedral unit. These units are connected via a flexible Si-O-Si bond, sensitive to vdW forces (see Fig. \ref{sipoly}). Cristobalite (space group P4$_1$2$_1$2) has a similar local structure and only 5\% larger volume. The energy difference 
between $\a$-quartz and cristobalite is found to be approximately 20 meV in experiment \cite{siocris1,siocris2}. When the ZPE contribution is subtracted the energy difference doubles \cite{haysio}. While LDA predicts the correct energy order PBE erroneously predicts cristobalite to be lower in energy due to missing vdW forces \cite{haysio}. A similar effect was recently observed in the borate (B$_2$O$_3$) polymorphs \cite{prm}. Stishovite (space group P4$_2$/mnm) is a high-pressure phase with 37\% higher density, having sixfold coordinated Si atoms. The energy difference between stishovite and $\a$-quartz is experimentally determined to 0.51-0.54 eV (see Ref. \cite{sio2pbe} and experimental references therein \cite{sio2exp1,sio2exp2}). LDA underestimates this value by more than 0.4 eV, whereas PBE here predicts the correct energy difference \cite{sio2pbe}. 

In the lower panels of Fig. \ref{sipoly1} we illustrate the results for the $\a$-quartz-cristobalite (red) 
and $\a$-quartz-stishovite (yellow) energy differences. We optimized all structures with PBE plus a vdW correction at the Tkatchenko-Scheffler (TS) 
level \cite{vdwts}. 
In the case of $\a$-quartz-cristobalite, LDA gives a value of 33 meV and PBE0 a value of 12 meV. As discussed in Ref. \cite{haysio} a vdW correction improves this result. 
RPA gives a value of 31 meV, which is underestimated by 15 meV according to QMC calculations. As the Si-O bond remains similar in $\a$-quartz and cristobalite, we interpret this relatively small error coming from the underestimated 
vdW forces with RPA. RPAx and SOSEX bring the result in good agreement with the QMC value of 43 meV. 

The situation is different in the case of stishovite which has sixfold coordinated Si atoms. Here the variations between the different methods are an order of magnitude larger. Subtracting the ZPE, calculated in Ref. \cite{haysio}, QMC predicts an energy difference of 0.51 eV \cite{qmcsio}, similar to the PBE0 result (0.52 eV) and experiment. With RPAx we find a value of 0.49 eV which is 80 meV larger than the RPA result (0.41 eV).  Although still somewhat underestimated we see that the exchange kernel is responsible for the improved accuracy. Self-consistency could further improve this result. That an xc kernel can improve the $\a$-quartz-stishovite energy difference was also found in Ref. \cite{ruzrpa} using the renormalized PBE kernel (0.51-0.54 eV). 

Our RPA result differs from the result presented in Ref. \cite{perdewsio} by 20 meV. The origin of this difference, occurring already at the PBE level, can be traced to the use of PAW (projected augmented wave) potentials \cite{paw} in Ref. \cite{perdewsio}. A larger difference can be found with respect to the RPA result presented in Ref. \cite{ruzrpa}, a difference that can be traced mainly to the EXX part of the energy (see Appendix C).

From the technical point of view the situation is similar to the BN polymorphs. In the upper panel of Fig. \ref{sipoly1} we show the convergence of the $\a$-quartz-stishovite correlation energy difference ($\D E_{\rm c}^{\rm q-s}$) with respect to $N_\nu$. Again, we find a slower convergence in RPA as compared to RPAx.
The RPA and RPAx values were converged within 10 meV using $30\times N_e$ and $10\times N_e$ eigenvalues, respectively.
\begin{figure}[t]
\includegraphics[scale=0.28,angle=90]{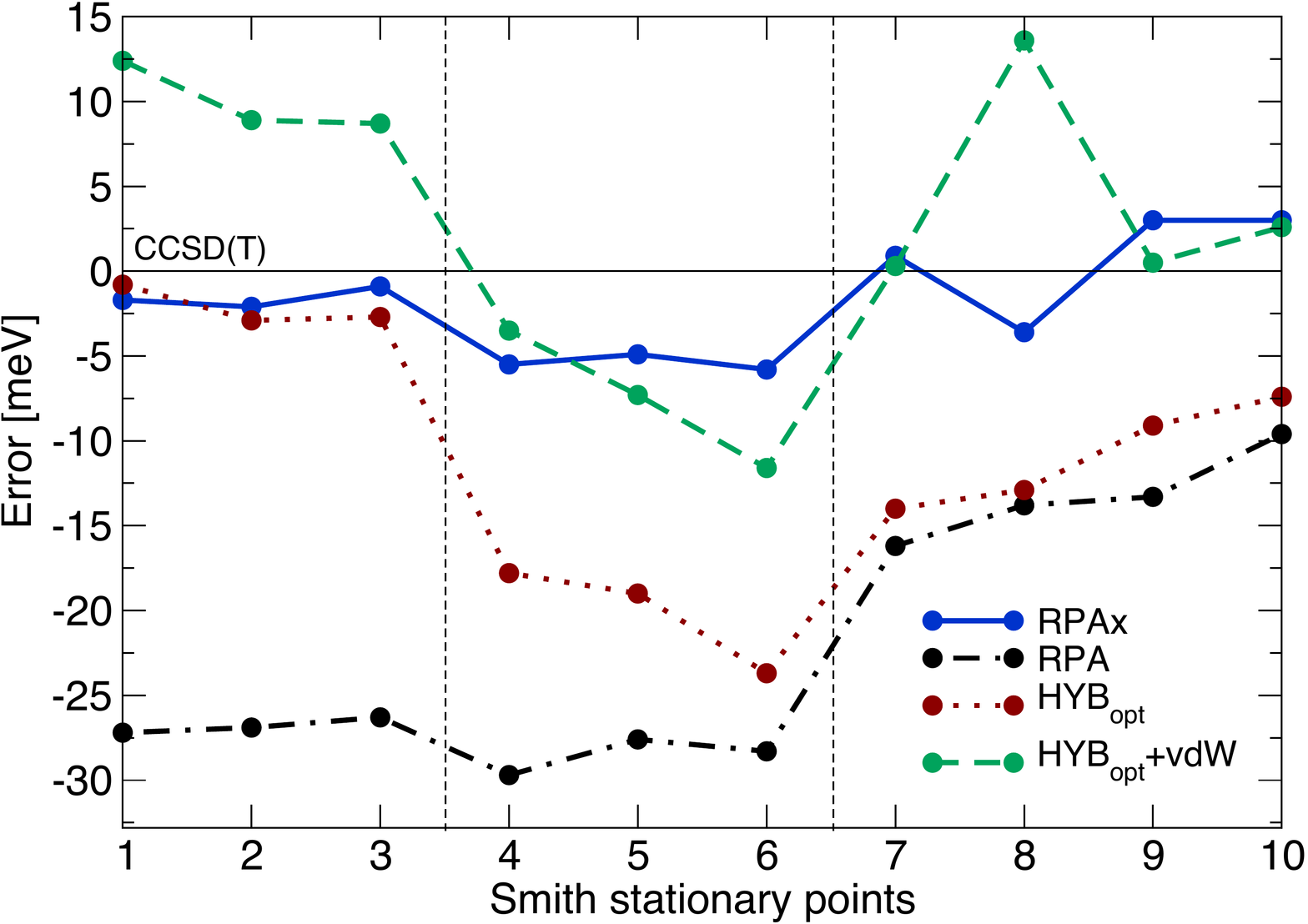}
\caption{Error in the binding energy (meV) with respect to CCSD(T) for the 10 Smith stationary points.  }
\label{spbind}
\end{figure}
\begin{figure*}[t]
\includegraphics[scale=0.28,angle=90]{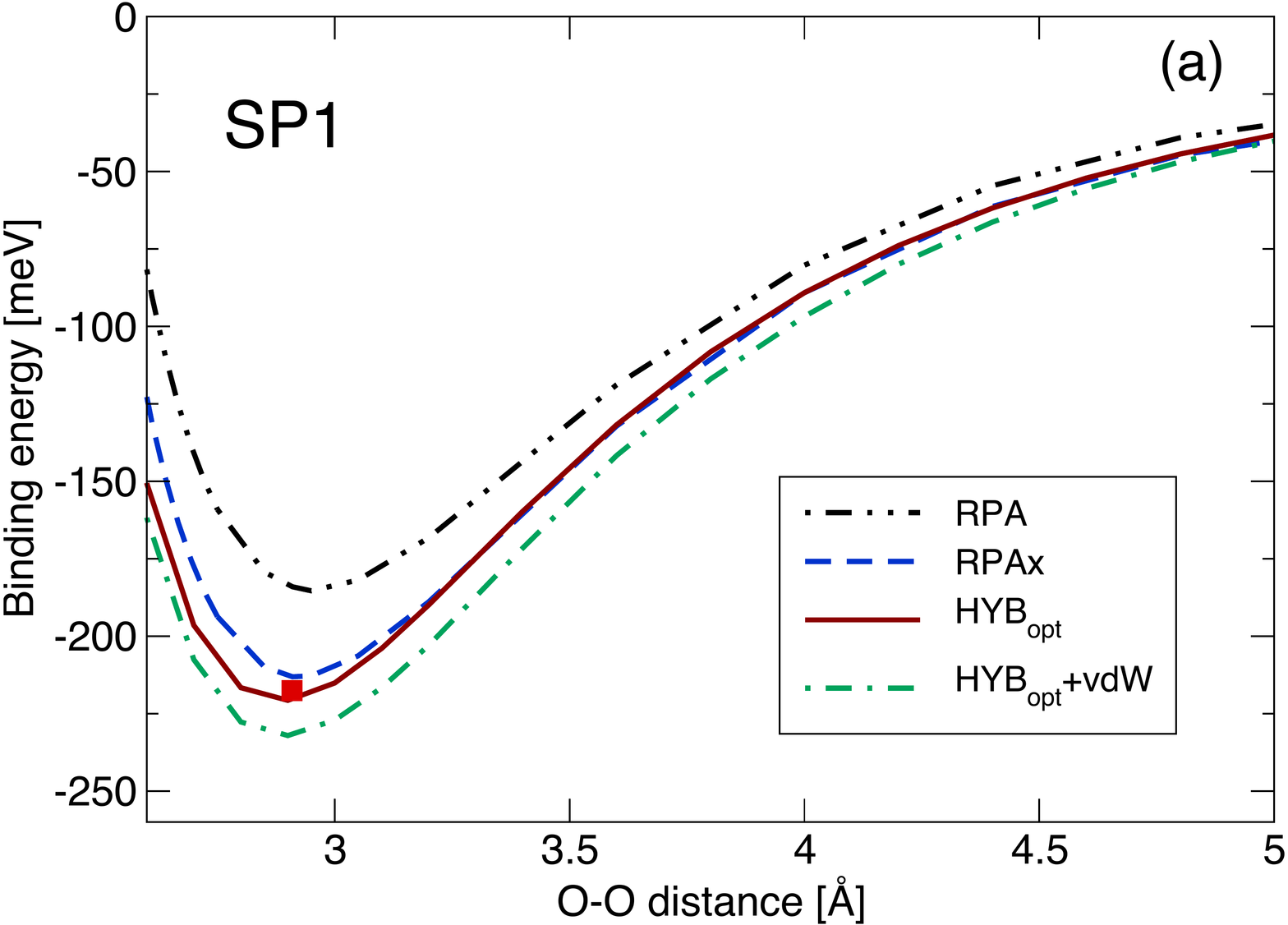}\hspace{5mm}
\includegraphics[scale=0.28,angle=90]{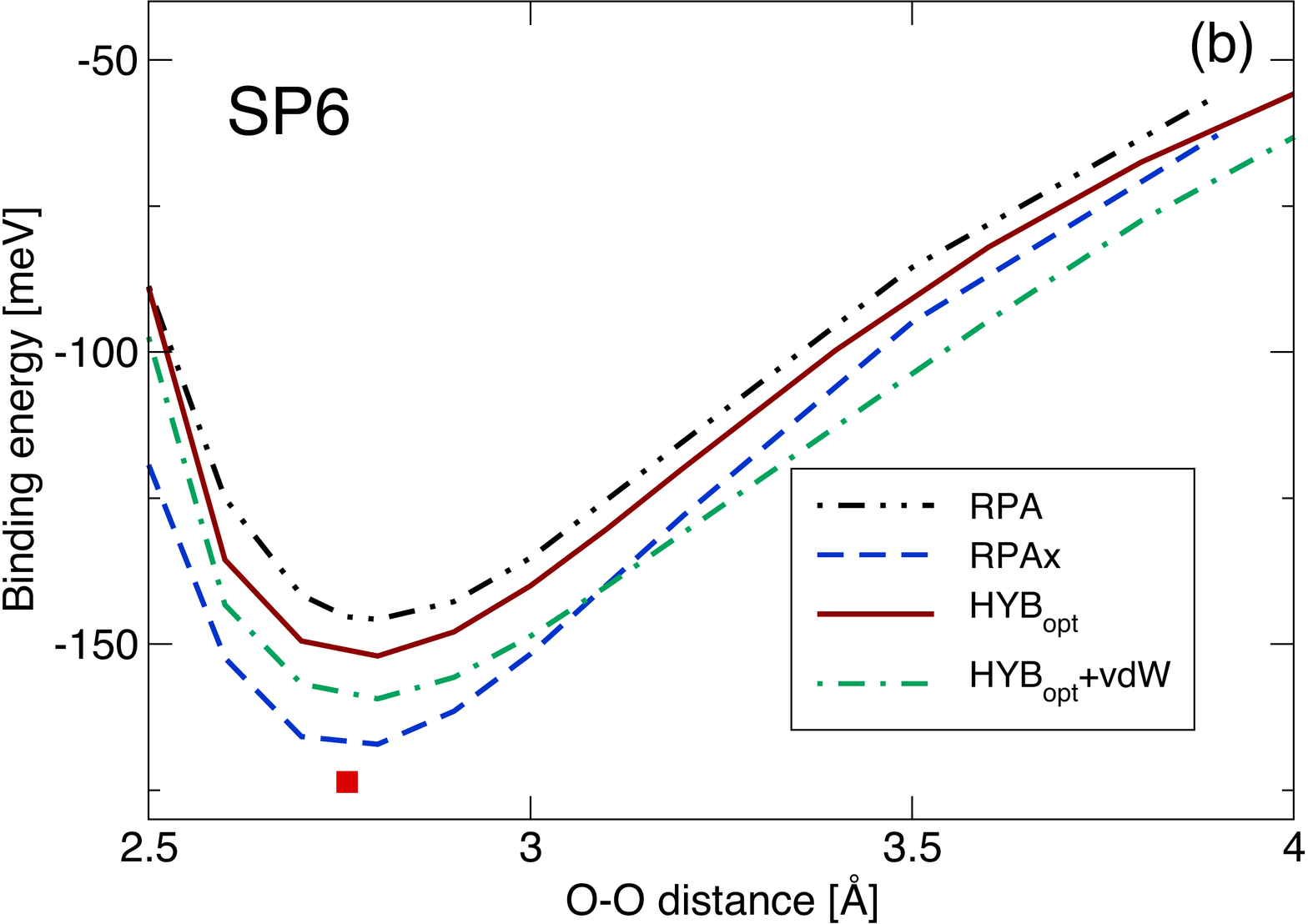}
\caption{Binding energy (meV) as a function of O-O distance (\AA) of water dimers in the SP1 (a) and SP6 (b) configuration. Red square corresponds to the CCSD(T) result \cite{ccsdtSP}.}
\label{dimer}
\end{figure*}
\section{Ice polymorphism}
The problem in predicting the relative energies between high- and low-pressure phases is relevant also for molecular solids 
such as the ice polymorphs. At ambient pressure the oxygen atoms in water form an 
hexagonal lattice held together by hydrogen bonds. By applying pressure the average 
O-O distance reduces such that the oxygen atoms are approached in non-hydrogen bonded configurations, which increases the coordination 
number from four up to eight in some dense high-pressure structures \cite{waterreview}. 
\begin{table}[b]
\begin{center}
\caption{Anisotropic polarizability (a.u.) of the water molecule. In the first two columns results are obtained 
with PBE orbitals and in the second two columns with the optimized hybrid orbitals (20\% of exchange 
for RPA and 35\% for RPAx). 
Geometry and CC results are from Refs. \onlinecite{ccsda,ccsdta}.}
\begin{ruledtabular}
\begin{tabular}{l| c c| c c| c c}
 & RPA &  RPAx  &  RPA & RPAx & CCSD &CCSD(T)   \\
 &\multicolumn{2}{l|}{PBE orbitals}&\multicolumn{2}{l|}{HYB$_{\rm opt}$ orbitals}& &\\
\hline
$\a_{xx}$       	 & 9.14  & 9.66  & 8.49  & 8.76 & 8.976 & 9.250 \\
$\a_{yy}$            & 9.19  & 10.35 & 8.78  & 9.69 & 9.712 & 9.874 \\
$\a_{zz}$       	 & 9.19  & 9.87  & 8.66  & 9.18 & 9.310 & 9.529 \\
$\a_{\rm iso}$       	 & 9.17 & 9.96  & 8.64  & 9.21& 9.33 & 9.55\\
\hline
MAE&-0.38 &+0.41 &-0.91 &-0.35 & -0.22&\\
\end{tabular}
\end{ruledtabular}
  \end{center}
  \label{polwater1}
\end {table}
Calculating the energy difference between these phases with standard functionals in DFT, again produces large errors. One of the failures has been attributed to the lack of vdW forces, which are expected to play an important role in high-pressure phases as well as liquid water \cite{watervdw,watervdw3,icets,icets2}. In addition, a fraction of HF exchange 
has shown to be important for the description of the hydrogen bond \cite{icets,icets2}. However, combining a hybrid functional 
with a vdW correction still does not give a complete description. The strongly constrained and 
appropriately normed (SCAN) meta-GGA functional \cite{scan_org} has shown that it is possible to obtain good energy differences between 
low- and high-density phases using a semi-local functional. 
On the other hand, lattice energies, i.e., the energy gain per monomer in the solid, remain overestimated with SCAN \cite{scan_nature}. 
The more advanced RPA also provides improved energy differences, but in this case with underestimated lattice energies \cite{kresseice}, due to the poor description of the hydrogen bond within RPA \cite{rparangewater,rocca,hcdg18}. Calculations using second order M\o ller-Plesset perturbation theory have, indeed, shown that higher order exchange terms are important \cite{ccsdtice}. 

We will now investigate the performance of the RPAx, and start by exposing a 
fundamental problem of the RPA when calculating the polarizability of the water molecule. 
Since the static polarizability $\a(0)$ (Eq. (\ref{isoa})) is not a variational quantity like the total energy, it has a stronger dependency on the input orbitals (and eigenvalues) \cite{gisbergen}. 
It can be argued that the correct orbitals should be those coming from a self-consistent RPA/RPAx calculation \cite{hvb10}. 
However, instead of these orbitals, we will use both PBE orbitals and orbitals from an optimal 
hybrid functional based on a local KS potential \cite{tise2_prb}. The optimization of the hybrid functional is carried out by 
minimizing the total RPA/RPAx energy with respect to the fraction of exact-exchange in hybrid functional used to generate the input orbitals (see further details in Appendix A and Ref. \cite{oepdg}). 
In order to study effects of anisotropy we calculate all diagonal elements 
of the polarizability. The results, obtained using the geometry of Refs. \onlinecite{ccsda,ccsdta}, are presented in Table I. We see that 
RPA clearly underestimates the polarizability and the magnitude of anisotropy when evaluated with optimal hybrid orbitals 
(20\% of exchange in RPA). The isotropic RPA polarizability improves with PBE orbitals, but the anisotropy is then completely lost. 
The origin of any anisotropy within RPA is, thus, solely due to an effect of exchange in 
the description of the orbitals. In contrast, the RPAx exhibits anisotropy already with PBE orbitals stemming 
from the exchange kernel. This effect is enhanced using optimal hybrid orbitals (35\% of exchange in RPAx). 
The RPAx agrees very well with CCSD results \cite{ccsda}, and are not so far from CCSD(T) results \cite{ccsdta}. 
These results highlight the important role of the exchange kernel for describing not only the water molecule, but also 
the long-range interaction between water molecules. 

In order to study the water intermolecular interaction in more detail we determine the binding 
energy of the ten Smith stationary points (SPs) \cite{smithp}. In addition to the global minimum, the SPs 
correspond to transition states or higher order saddle points on the water dimer potential energy surface \cite{ccsdtSP}. The geometries of the 
SPs, depicted Appendix C are relevant for liquid water \cite{waterbench,qmccluster}, and some are found in 
high-pressure ice polymorphs. 
The lowest energy configuration (SP1) corresponds to the optimal hydrogen bond geometry, included in many 
molecular test-sets. Previous work has shown that RPAx and related methods capture the binding energy of 
this configuration rather well, while RPA fails by around 15\% \cite{rparangewater,rocca,hcdg18}. 
In Fig. \ref{spbind} we have summarized the results for all ten SPs using structures optimized at the 
CCSD(T) level \cite{ccsdtSP}. The HYB$_{\rm opt}$ 
corresponds to the hybrid functional optimised with RPAx (35\% of exchange). The HYB$_{\rm opt}$ orbitals were used for evaluating the energy in both RPA and RPAx. For the total energy the choice of orbitals has a small effect. The largest difference is found for the SP1 configuration where the RPAx binding energy change from 212 meV to 216 meV when going from PBE to HYB$_{\rm opt}$ orbitals. We have also included a vdW correction 
at the TS level in combination with the same hybrid (HYB$_{\rm opt}$+vdW). 
Errors are determined with respect CCSD(T) reference values \cite{ccsdtSP}. It is clear that RPA fails not only for SP1 
but for all configurations. The error is, however, rather constant suggesting that the energy differences are still quite well described. The hybrid functional performs very well for 
the first three hydrogen bonded configurations but fails, similarly to RPA, for the others. Energy differences 
will, in this case, not benefit from error cancellation. Adding the TS-vdW correction shifts all values with a 
similar constant (except for SP8). The vdW correction thus improves results for some configurations and worsens the results for others, but 
the quality of the energy difference will be similar to that of the hybrid functional without a vdW correction. 
These results show how complex and challenging it is to capture the full (H$_2$O)$_2$ potential energy surface. 
RPAx is the only approximation that performs well for all SPs and produces results comparable to reference 
values at the CCSD(T) level. 
\begin{figure*}[t]
\includegraphics[scale=0.47]{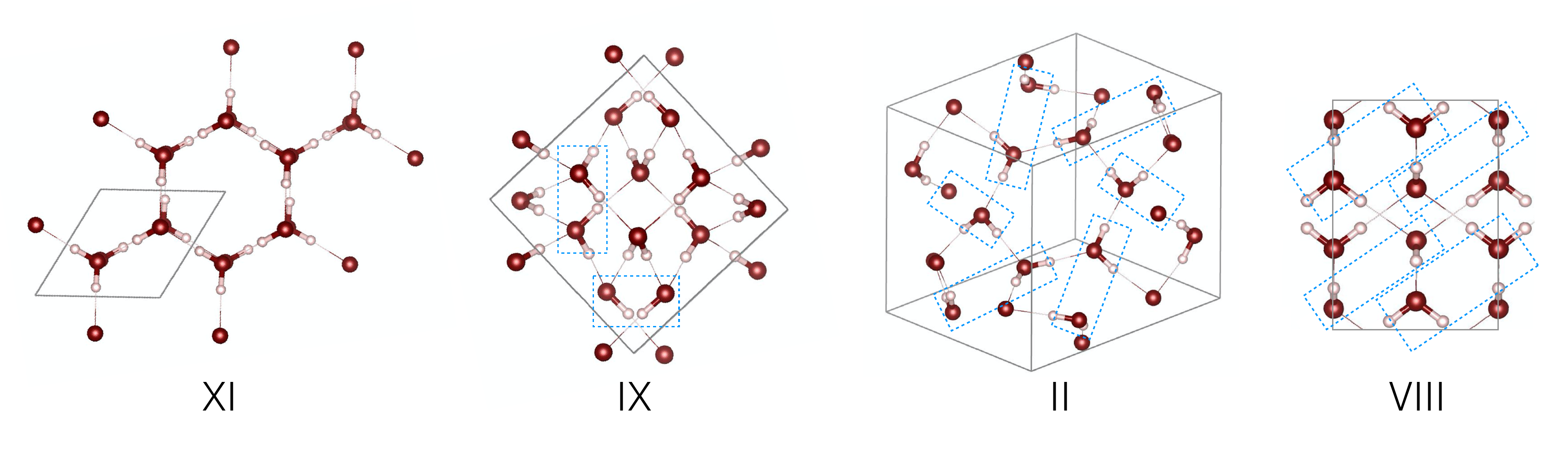}
\caption{Structures of the four ice polymorphs studied in this work (XI, IX, II, VIII). Blue dashed boxes indicate water 
dimers in non-hydrogen bonded configurations but with comparable O-O distance.}
\label{ice}
\end{figure*}

In Fig.~\ref{dimer} we plot the dissociation energy curves of the SP1 and SP6 configurations. 
The dimers are stretched with fixed intramolecular geometries. Given the accuracy of the RPAx we can 
now use this approximation as a reference to compare the other methods too. It is interesting to note that the 
hybrid functional which gives a good value at the SP1 bond midpoint also captures the dissociation tail. 
RPAx and HYB$_{\rm opt}$ perfectly coincides up to 5 \AA. Adding a vdW correction improves the binding 
energy of the SP6 configuration. However, from the full dissociation curve, we see that this is at the expense 
of producing the wrong tail. Even in this non-hydrogen bonded configuration, RPAx and HYB$_{\rm opt}$ are closer
at stretched geometries. 

We are now ready to study solid forms of water. 
In Fig.~\ref{ice} we illustrate four different phases of ice in the order of decreasing volume per molecular 
unit: XI, II, IX and VIII. All structures are relaxed with the PBE+TS functional. Due to the smaller unit cell, we have chosen the 
ferroelectrically proton-ordered ice XI (Cmc2$_1$) instead of the ordinary disordered I$_{h}$ \cite{icexiexp,icexistruc}. 
The energy of XI is three meV lower than I$_{h}$ in RPA \cite{kresseice}. Both are purely hydrogen bonded solids 
with water molecules arranged in SP(1-3)-like configurations. In clusters and solids the strength of the hydrogen bond is enhanced as compared 
to the isolated dimer bond strength. This effect is captured by most functionals. In Table II we present lattice energies using HYB$_{\rm opt}$, HYB$_{\rm opt}$+vdW, RPA, RPAx and SOSEX. The results are compared 
to CCSD(T) \cite{ccsdtice} and recent QMC results \cite{qmcice}. We see that both HYB$_{\rm opt}$ and RPAx produce a 
result for XI of similar quality as for the SP1 dimer. In the same way, RPA underestimates the lattice energy. 
We note that for the solids we have evaluated both RPA and RPAx with PBE orbitals in order to compare with 
previous RPA results \cite{kresseice}. The use of HYB$_{\rm opt}$ orbitals (always via a local potential) 
increases the XI result by 10 meV ($< 2\%$). 

\begin{table*}[t]
\caption{Lattice energies (meV) of ice XI, II, IX and VIII with the optimized hybrid functional HYB$_{\rm opt}$ (35\% exchange), the same hybrid functional with a TS-vdW correction, RPA, RPAx, SOSEX, CCSD(T) \cite{ccsdtice}, QMC \cite{qmcice} and experiment \cite{expice1}. The relative energies with respect to XI are also presented. For CCSD(T) and QMC XI should be replaced by I$_h$.}
\begin{ruledtabular}
\begin{tabular}{l r r r r r r r r}
  Ice &HYB$_{\rm opt}$& HYB$_{\rm opt}$+vdW& RPA & RPAx & SOSEX &CCSD(T)& QMC &Exp.  \\\hline
XI (Cmc2$_1$)   &614 &676&551&609&634 &601 &615&610\\
II  &559&671&542& 609&639&601 &613&609\\
IX   &570 &673&538 &598&625&- &- &606\\
VIII &471& 606&516&584& 613&574 &594&595\\\hline
II\textemdash XI  &55& 5 &9&0&-5& 0 & 2&1\\
IX\textemdash XI &44& 3 &13&11&9 &- &- &4\\
VIII\textemdash XI&143&70&35& 25&21&27 &21&15\\
\end{tabular}
\end{ruledtabular}
\end {table*}

Ice IX and II have similar volumes, approximately 20\% smaller than XI.
The smaller volume makes non-hydrogen bonded configurations more relevant. In Fig. \ref{ice} we have identified pairs of H$_2$O monomers for which the magnitude of the O-O distance is approaching that of hydrogen-bonded configurations. The ability of a functional to capture the energies of different dimer orientations thus increases in importance for the IX and II phases. In Table II we see that the good performance of HYB$_{\rm opt}$ is now lost, producing underestimated lattice energies. Adding the vdW correction overcorrects the lattice energies but recovers the near degeneracies between XI, II and IX. This result was highlighted in Ref. \cite{icets} suggesting that the vdW forces play an important role in the II and IX phases. We note that the most recent QMC calculation predicts ice II and I$_h$ to be nearly  degenerate at 613 and 615 meV, respectively (with a 5 meV error bar) \cite{qmcice}, while an earlier QMC calculation from 2011 predicts ice II to be lower in energy than I$_h$ by 4 meV (609 and 605 meV respectively) \cite{icets}. The corresponding experimental results have been extracted in Ref. \cite{expice1}, yielding 609 and 610 meV respectively. Our RPA calculation gives an II-XI energy difference of 9 meV, in good agreement with previous calculations \cite{kresseice}. RPAx clearly reduces this energy difference making them degenerate. SOSEX lowers the energy of ice II with respect XI even further, reverting the energy order. Given that ice XI is expected to be a couple of meV lower in energy than I$_h$, even by experiment \cite{expice2}, our results indicate that the II-I$_h$ energy difference should be $\leq$ 0 meV. 
We also note that HYB$_{\rm opt}$ and HYB$_{\rm opt}$+vdW predict ice IX to be lower in energy than ice II, in contrast to the present (beyond-)RPA methods and the previously mentioned SCAN functional \cite{scan_nature}.

Ice VIII has a volume 40\% smaller than ice IX and is stable above 2 GPa. It corresponds to a proton-ordered phase of VII \cite{viii1,viii2}. In this structure the O atoms are eightfold coordinated and it is easy to identify both SP6-like and SP8-like dimers, in addition to the hydrogen bonded configurations. As pointed out in Ref. \cite{waterreview} the smallest O-O distance is actually between monomers in non-hydrogen bonded configurations. The experimental result for the lattice energy was initially determined to 577 meV \cite{expice1}. Later, this has been revised to 595 meV \cite{qmcice}. Again HYB$_{\rm opt}$ fails, while HYB$_{\rm opt}$+vdW only slightly overcorrects in this case. The energy difference with respect to ice XI remains, however, largely overestimated. The error of the RPA is slightly larger than for the other polymorphs. RPAx is again very accurate lying between CCSD(T) and QMC. 

Overall the RPA results for the ice polymorphs are consistent with the analysis of the SPs above. Lattice energies are underestimated by approximately 13\%, which is very similar to the error found for the SPs. Relative energies are much better described, benefiting from error cancellation. RPAx, which captures the essential correlation effects, gets both lattice energies and relative energies in good agreement with experiment and more sophisticated methods. SOSEX slightly overestimates the hydrogen bond and thereby overestimates the lattice energies for all polymorphs.
\section{Conclusions}
The RPA is becoming an important tool in materials science for calculating ground-state properties of solids. There are, however, some inherent limitations of the RPA which can cause significant errors in certain systems or situations. In this work we have analyzed some of these cases and demonstrated how exact-exchange in the response function provides a theoretically well-defined, accurate and reliable improvement. 

In general, the exchange term increases atomic polarizabilities leading to, e.g., improved vdW coefficients. Here we have shown that for the cohesive energy of purely vdW bonded solids, such as Ar and Kr, RPAx produces very accurate results, improving the RPA by more than 20\%. vdW forces also play an important role in the SiO$_2$ and BN polymorphs. We have shown that the interlayer binding energy of r-BN is enhanced with RPAx. However, with a smaller correction than for purely vdW bonded systems. The $\a$-quartz-cristobalite energy difference is also enhanced with RPAx, and now agrees with reference QMC calculations.

Earlier work have shown that the RPAx corrects the overestimated RPA correlation energy. As often pointed out, this error cancels to a large extent when looking at energy differences of similar systems. Which systems are similar enough is, however, not well-defined. In this work we have highlighted that when the coordination number changes, as between 
high- and low-pressure phases, errors do not cancel to a satisfactory degree. This has been demonstrated for the r-BN-c-BN and the $\a$-quartz-stishovite energy differences. In the first 
case, RPA and RPAx predict different energy ordering, and in the second case, RPAx strongly enhances the energy difference. In both cases RPAx produces results in line with highly accurate methods such as CC or QMC. 

We have also studied the (H$_2$O)$_2$ potential energy surface and ice polymorphism. The errors of RPA (and hybrid functionals) in describing the lattice energies of ice IX, II, XI and VIII were analysed in terms of the polarizability of the water molecule and the 10 SPs. 
The RPA clearly underestimates both magnitude and anisotropy of the polarizability, but captures rather well the energy differences between different H$_2$O dimer configurations thanks to a good error cancellation. This implies underestimated lattice energies of ice but good relative energies. The RPAx, which includes higher order exchange effects, captures the correct strength of the hydrogen bond and is able to describe the full anisotropic interaction between H$_2$O monomers. Not only relative energies but also lattice energies are, thereby, in very good agreement with experiment and more sophisticated methods. 

Regarding the computational cost and feasibility of the RPAx the main constraint is the $N^3_k$ scaling, which can make k-point demanding systems several times more expensive than RPA. On the other hand, the convergence with respect to eigenvalues in the response function is in many cases faster with RPAx. Other parameters such as the total number of k-points or frequency points have a similar behaviour in RPA and RPAx. For the systems studied here, the cost of the RPAx calculation is 3-10 times that of the RPA calculation. 

All calculations were done non-self-consistently, on top of a PBE ground-state. For the water dimers and ice we showed that the use of hybrid orbitals, which can be considered as a step towards self-consistency, increases the energy differences by a few meV only. Further developments in this direction are, however, interesting for eventually calculating forces and lattice vibrations within the same framework. 

To conclude, we have shown that the RPAx opens up for highly accurate density functional calculations on solids that can address structural phase transitions involving distinct but nearly degenerate phases, and provide more reliable reference results than RPA when experimental data are missing or difficult to access.
\\
%
\acknowledgements
The authors would like to thank Lorenzo Paulatto, Guillaume Ferlat and Michele Casula for discussions. 
The work was performed using HPC resources from GENCI-TGCC/CINES/IDRIS (Grant No. A0050907625). 
Financial support from Emergence-Ville de Paris is acknowledged.
\appendix
\newpage
\section{Exchange parameter for water}
\begin{figure}[t]
\includegraphics[scale=0.32]{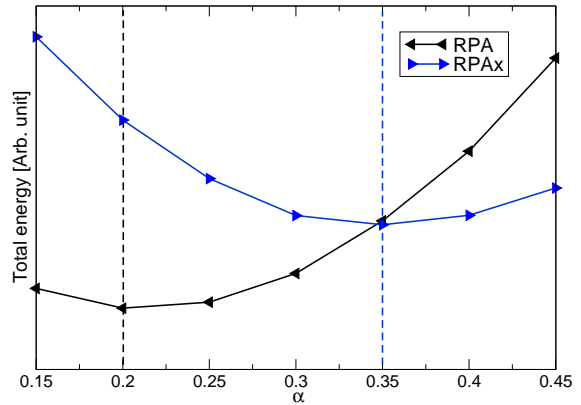}
\caption{The $\a$-parameter of the hybrid functional was chosen so as to minimize the total energy of the water molecule. The minimum occurs at 20\% in RPA and at 35\% in RPAx.}
\label{opta}
\end{figure}

The polarizability is more sensitive to the choice of input orbitals than the total energy. In Ref. \cite{hvb10} the atomic polarizabilites were evaluated with self-consistent RPA/RPAx orbitals. Since self-consistent RPA/RPAx is not yet available for molecules we have, in this work, evaluated the polarizability with PBE orbitals and orbitals coming from PBE0 hybrid functional, in which the $\a$-parameter was optimized by minimizing the total RPA/RPAx energy (see Fig. \ref{opta}). 

The local xc potential of the hybrid functional is defined as the functional derivative of the hybrid xc energy with respect to the density. It can thus be decomposed into
\be
v_\xc^{\rm hyb,\a}=v^{\a}_{\x}+(1-\a)v^{\rm PBE}_{\x}+v^{\rm PBE}_{\rm c}.
\label{lhyb}
\ee
For the exchange part, an integral equation known as the linearized Sham-Schl\"uter equation has to be solved
\be
\int\!d2\, \chi_s(1,2) v^{\a}_{\x}(2)=\a\int\!d2d3\,\L_s(3,2;1) \S_s^{\rm HF}(2,3)
\label{exxlss}
\ee
where $\S_s^{\rm HF}(2,3)$ is the Fock self-energy, $\L_s(3,2;1)=-iG_s(3,1)G_s(1,2)$ is the product of two Kohn-Sham Green's functions, $G_s$, and $\chi_s=\L_s(2,2;1)$. Details of the implementation can be found in Refs. \cite{oepdg,tise2_prb}

\section{Convergence of $N_\nu$ and $N_k$}
We include here additional results for the convergence of crucial parameters such as the number of eigenvalues in the response function ($N_\nu$), and the number of k-points ($N_k$). We used an energy cut-off of 80 Ry in all calculations except for the SiO$_2$ polymorphs where we converged the correlation energy with 60 Ry. 

\begin{figure*}[t]
\hspace{0.0cm}
\includegraphics[scale=0.3,angle=0]{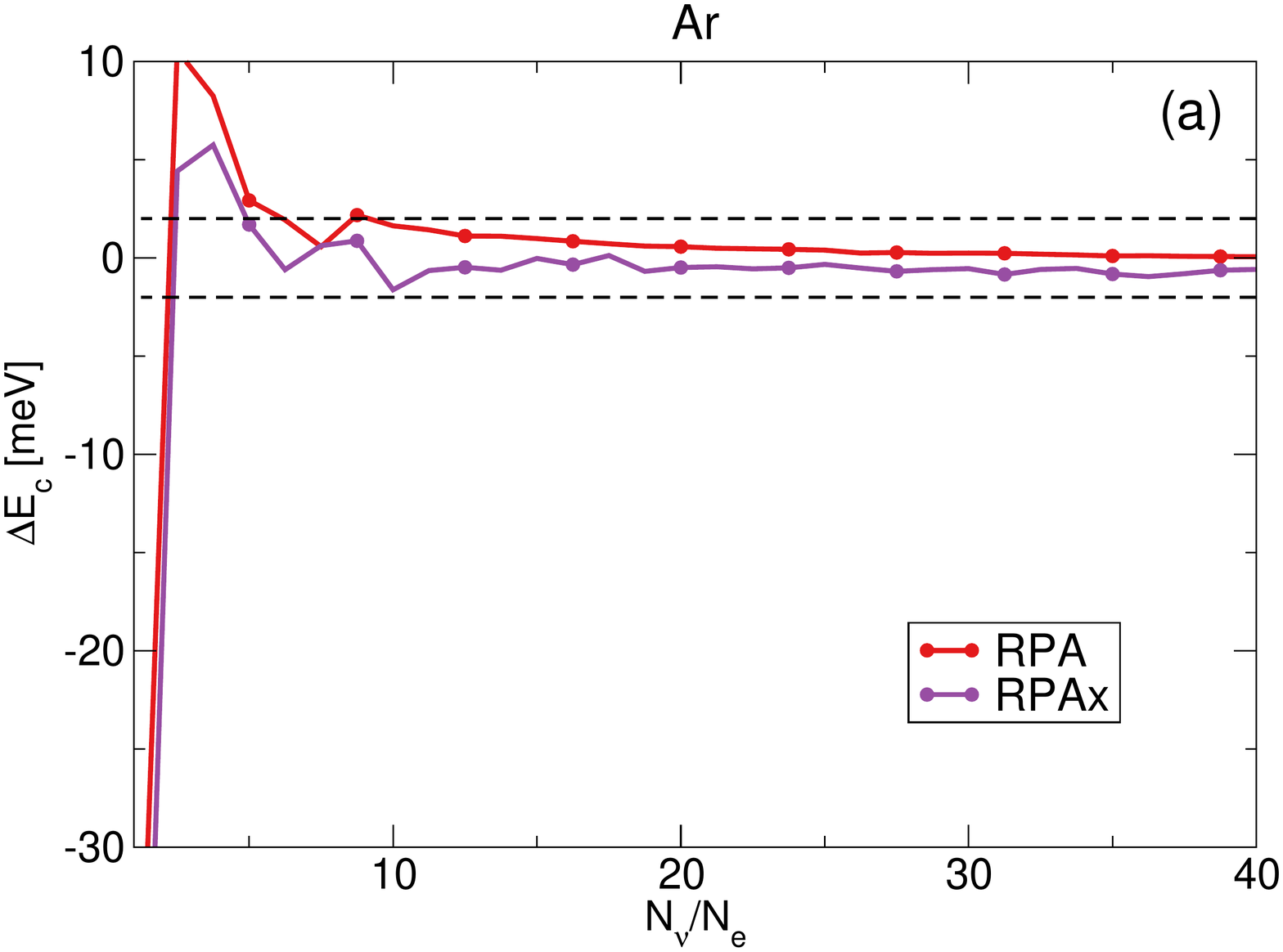}
\includegraphics[scale=0.3,angle=0]{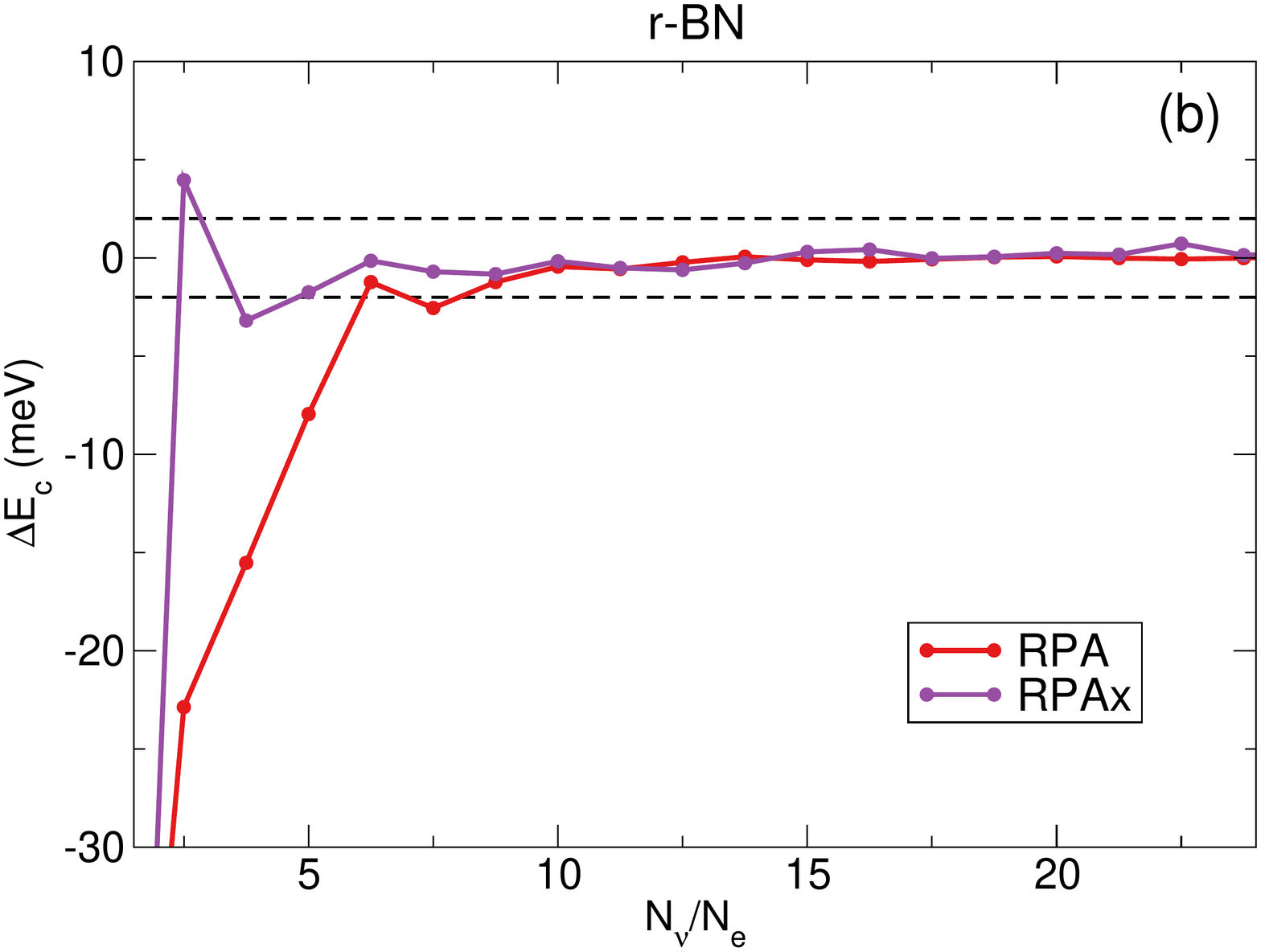}\\
\includegraphics[scale=0.3,angle=0]{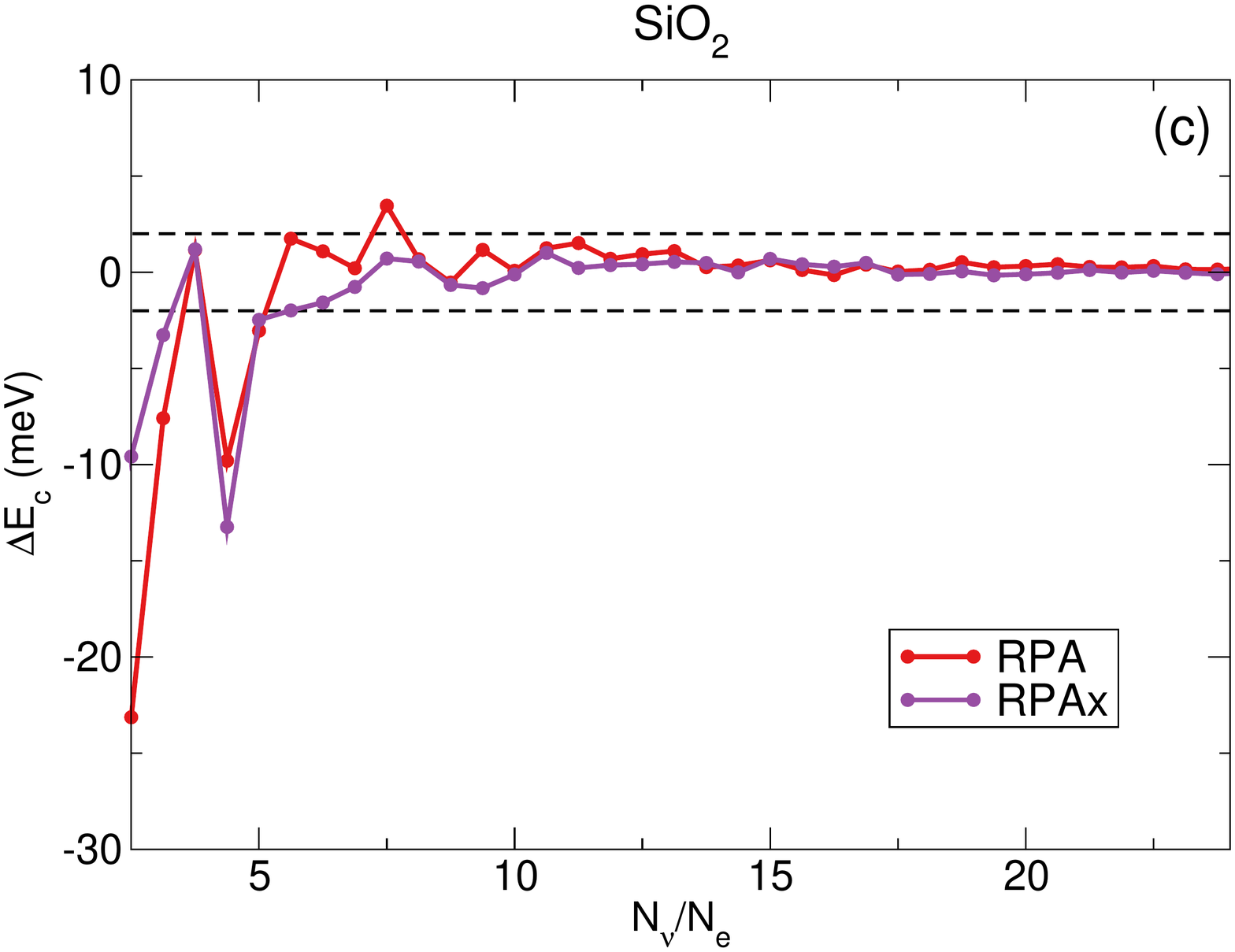}
\includegraphics[scale=0.3,angle=0]{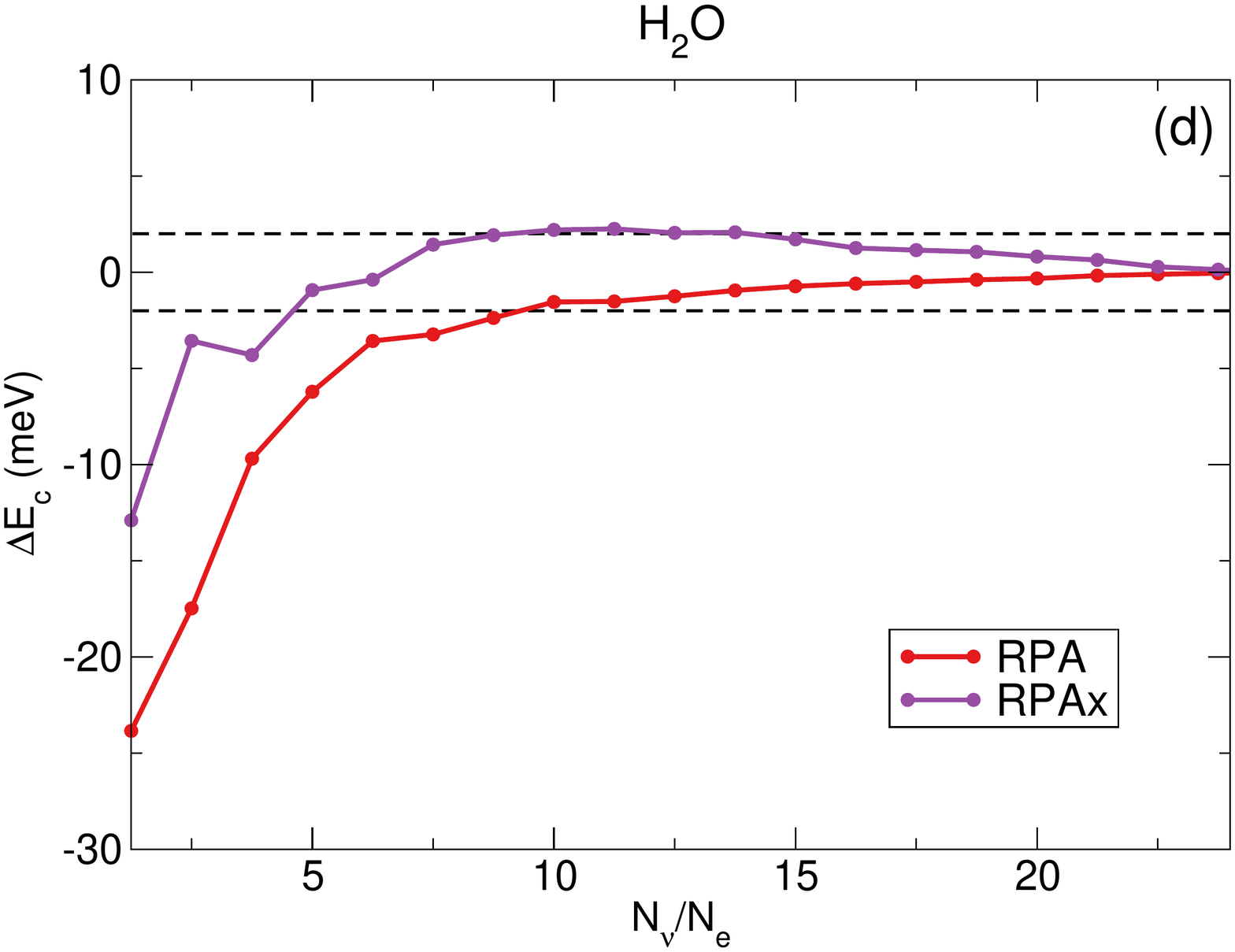}
\caption{Convergence of the correlation energy difference with respect to the number of eigenvalues per electron in the response function. (a) correlation energy difference between solid Argon and the isolated Ar atom, (b) between layered r-BN  and an isolated BN layer, (c) between $\a$-quartz and cristobalite and, (d) between the water dimer and an isolated water molecule.}
\label{neig}
\end{figure*}

The convergence with respect to $N_\nu$ for the $\a$-quartz-stishovite and c-BN-r-BN energy difference is presented in the main text. Here, in Fig. \ref{neig}, we present the $N_\nu$-convergence for the cohesive energy of Ar, the interlayer binding energy of r-BN, the $\a$-quartz-cristobalite energy difference and the binding energy of the water dimer. For these systems a value $N_\nu < 10*N_e$ is sufficient to converge within 2 meV for both RPA and RPAx. 

\begin{figure*}[t]
\includegraphics[scale=0.3,angle=0]{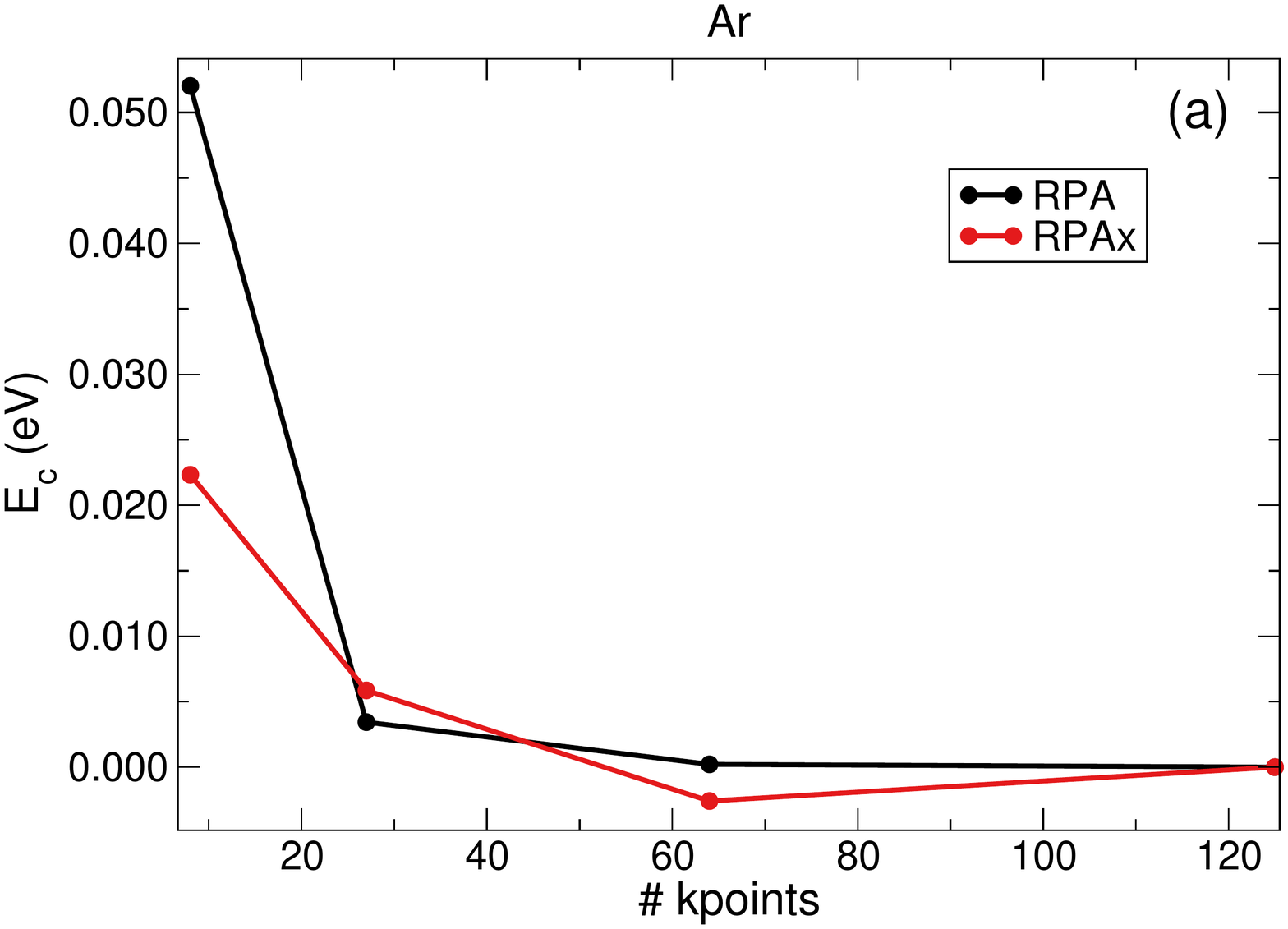}
\includegraphics[scale=0.3,angle=0]{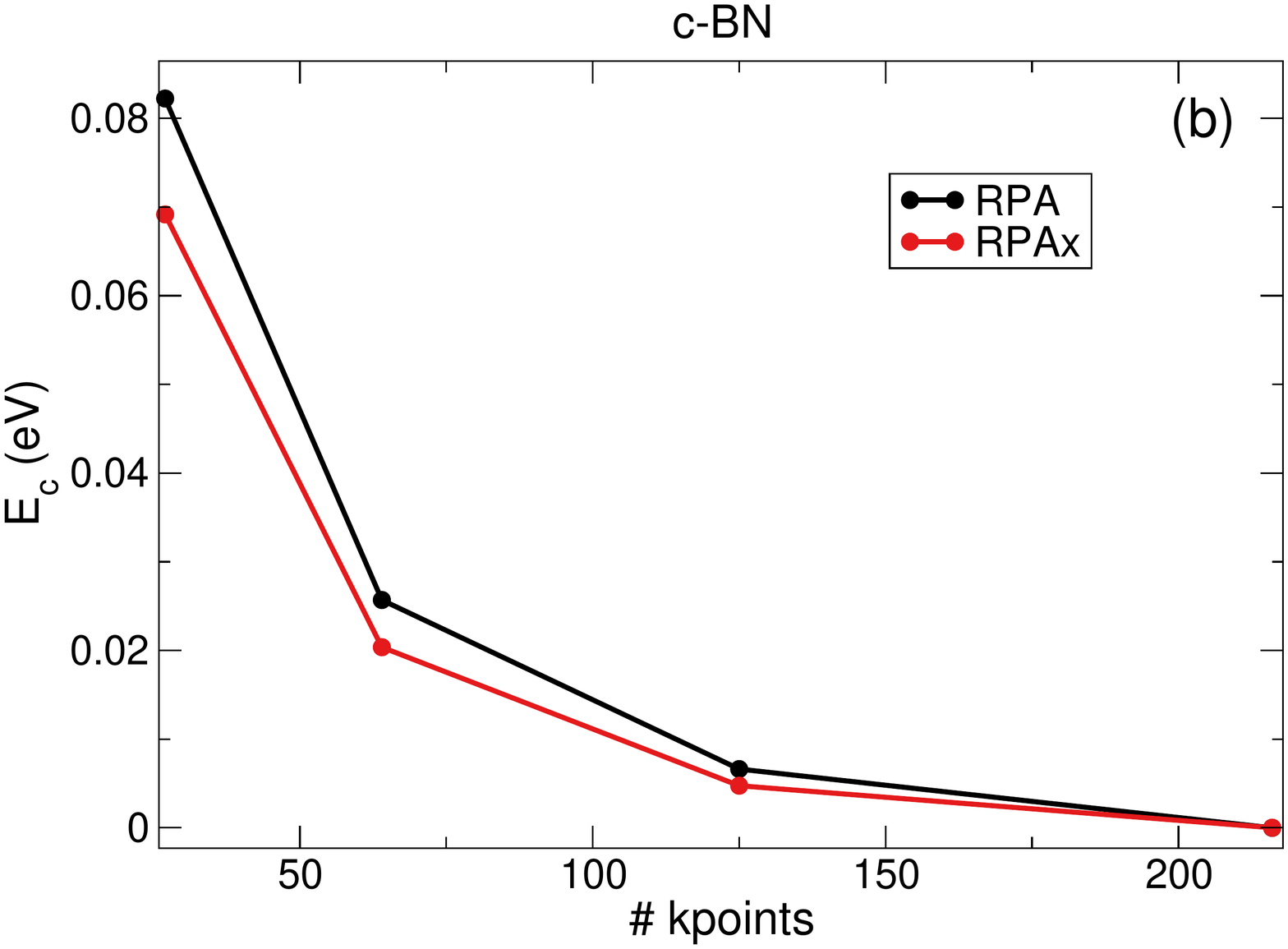}
\hspace{0.0cm}
\includegraphics[scale=0.3,angle=0]{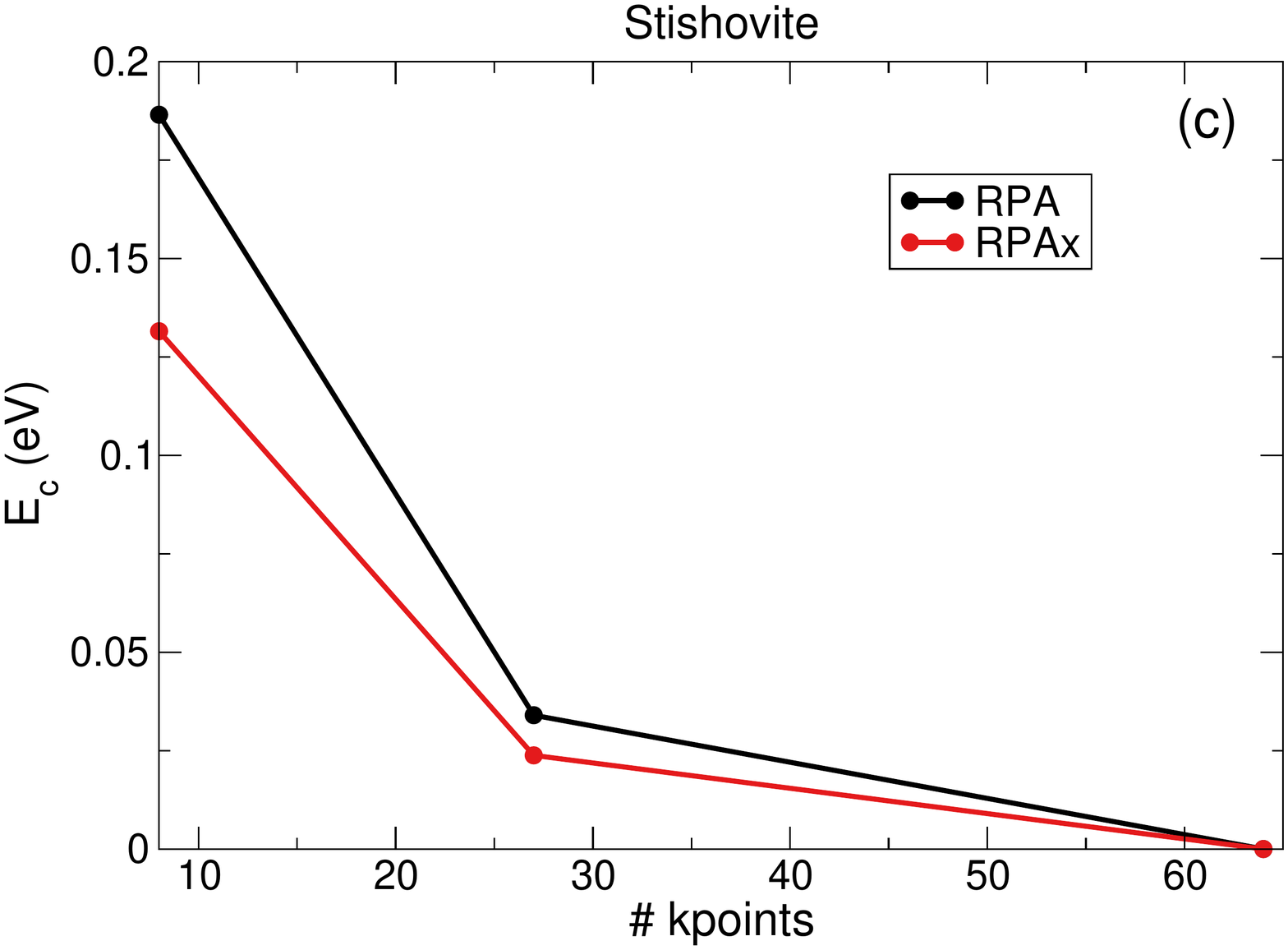}
\includegraphics[scale=0.3,angle=0]{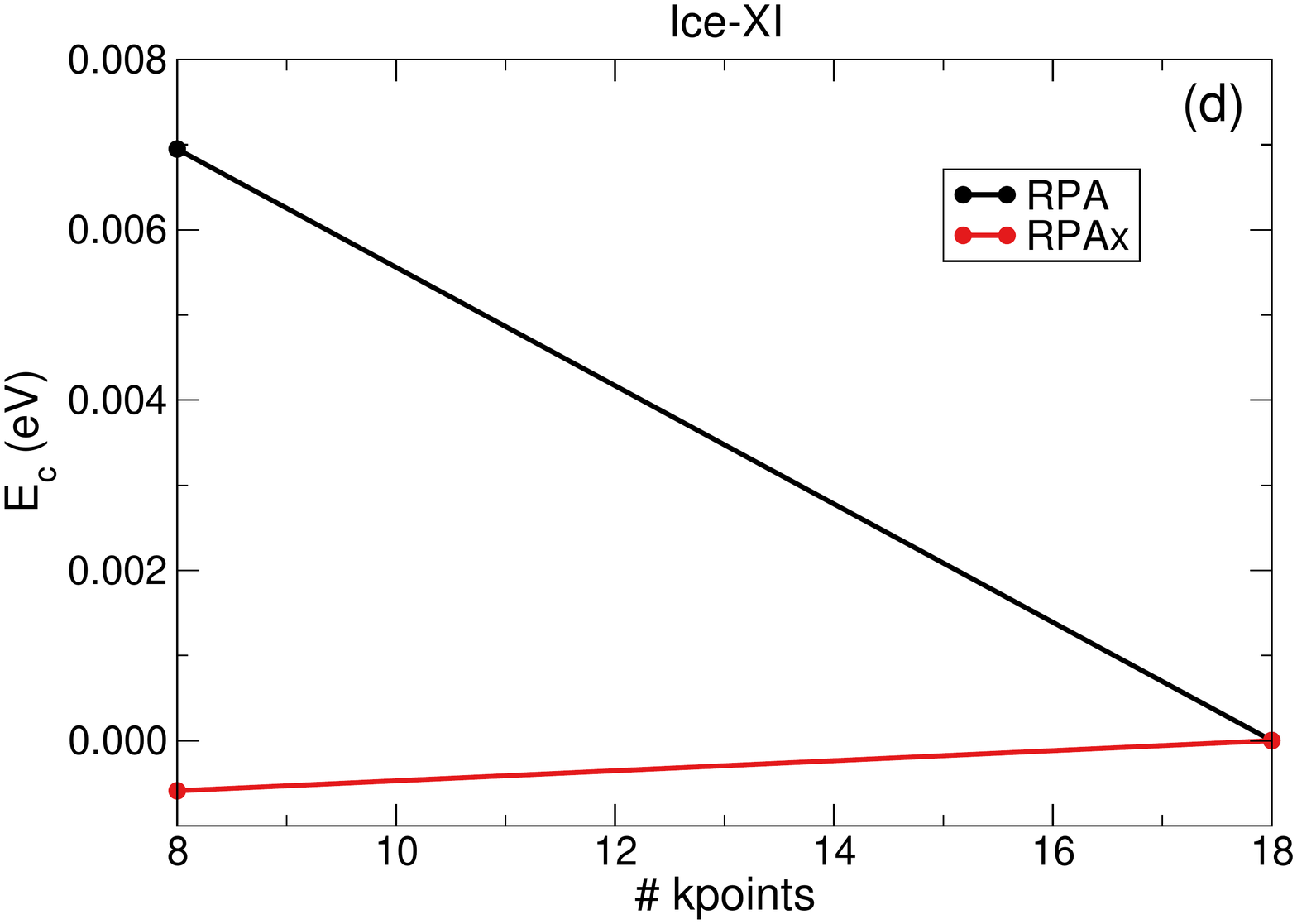}
\caption{Convergence of the correlation energy of (a) Ar, (b) c-BN, (c) stishovite and (d) Ice-XI with respect to the number of k-points. 
Curves have been shifted vertically by subtracting the value at the last point.}
\label{nkfig}
\end{figure*}

In Fig. \ref{nkfig} we present the convergence of the correlation energy with respect k-points for Ar, c-BN, stishovite and Ice-XI. We used shifted grids for faster convergence. Ar and Kr were converged within 2 meV using a $5\times 5\times 5$ grid. For c-BN the last point corresponds to a $6\times 6\times 6$ grid. For stishovite the last point corresponds to a $4\times 4\times 4$ grid. In order to save computing hours, the last RPAx point for stishovite is 
obtained by extrapolation, using the fact that the ratio between the RPA and RPAx correlation energy differences is essentially a constant. The last point for Ice-XI corresponds to a $3\times 3\times 2$ grid. 


\FloatBarrier

\section{Tables}
\begin{figure*}[t]
\includegraphics[scale=0.5]{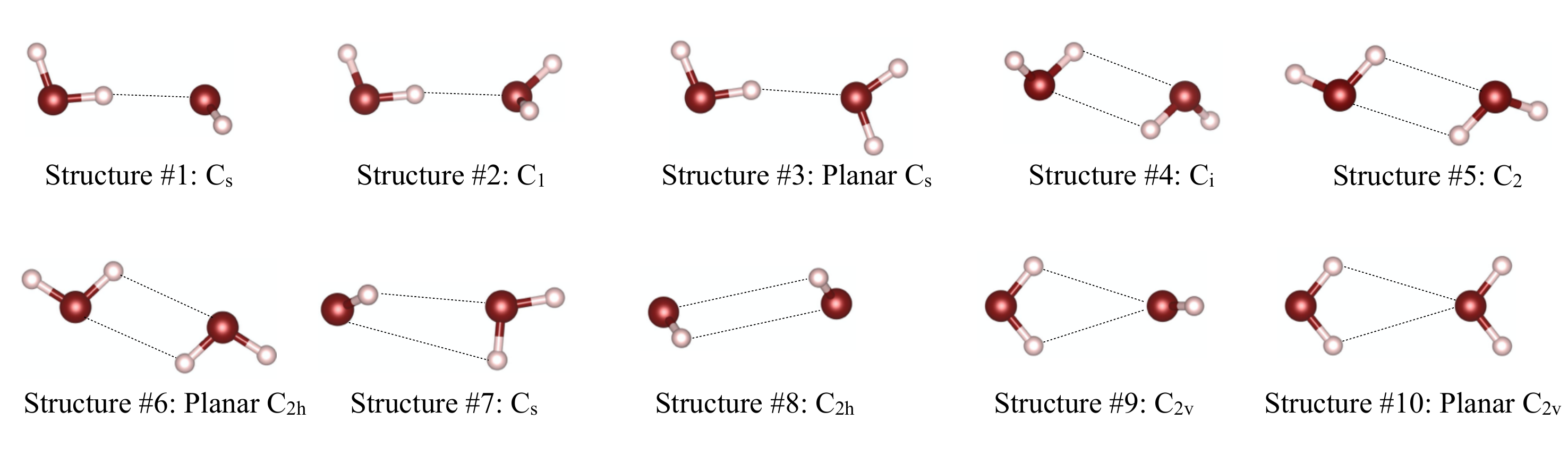}
\caption{Illustration of the 10 Smith stationary points of the (H$_2$O)$_2$ potential energy surface.}
\label{smith}
\end{figure*}

We have included 6 tables. Table \ref{tab1} contains the cohesive energy of Ar and Kr, and table \ref{tab2} the interlayer binding energy of r-BN. Table \ref{tab3} contains the energy of r-BN and w-BN with respect to c-BN, and table \ref{tab4} the energy of cristobalite and stishovite with respect to $\a$-quartz.  Tables \ref{tab5} and \ref{tab6} contain the binding- and relative energies of the 10 Smith points (SPs), respectively. An illustration of the 10 SPs can be found in Fig. \ref{smith}.

\begin{table}[ht]
\caption{\label{data} Cohesive energies of Ar and Kr (meV) in LDA, EXX, RPA, RPAx, SOSEX, CCSD(T) \cite{stoll99} and experiment \cite{aziz}.}
\begin{ruledtabular}
\begin{tabular}{l r r r r r r r}
  Solid  &LDA & EXX&RPA & RPAx & SOSEX &CCSD(T) & Exp. \\\hline
Ar  &141&-104&69&86 &91&83 &89 \\
Kr  &169&-102&94&117 &120& 114&123\\
\end{tabular}
\end{ruledtabular}
\label{tab1}
\end {table}
\begin{table}[h]
\caption{\label{data} Interlayer binding energy of r-BN (meV) in LDA, EXX, RPA, RPAx and SOSEX. No accurate reference value is available.}
\begin{ruledtabular}
\begin{tabular}{l r r r r r r}
  Solid  &LDA &EXX &RPA & RPAx & SOSEX   &\\\hline
r-BN & 56&-58 &74& 80  & 80  &  \\
\end{tabular}
\end{ruledtabular}
\label{tab2}
\end {table}

\begin{table*}[ht]
\caption{\label{data} Energy difference with respect to c-BN (meV) in LDA, PBE0, EXX, RPA, RPAx, SOSEX, CCSD \cite{ccsdtbn} and CCSD(T) \cite{ccsdtbn}. }
\begin{ruledtabular}
\begin{tabular}{l r r r r r r r r}
  Solid  &LDA&PBE0& EXX &RPA & RPAx & SOSEX &CCSD&CCSD(T)   \\
\hline
r-BN    &105 &-67 &-442&25 & -15 &-23 &-28&-4  \\
w-BN    &36 &37 & 48 &38 &37& 37&48 &44\\
\end{tabular}
\end{ruledtabular}
\label{tab3}
\end {table*}

\begin{table*}[ht]
\caption{\label{data} Energy difference with respect to $\a$-quartz in LDA, PBE0, EXX, RPA, RPAx, SOSEX, QMC \cite{qmcsio,haysio} and experiment \cite{siocris1,siocris2,sio2exp1,sio2exp2}.}
\begin{ruledtabular}
\begin{tabular}{l r r r r r r r r}
  Solid  &LDA&PBE0& EXX&RPA & RPAx & SOSEX &QMC & Exp.  \\
\hline
Cristobalite  (meV)  &32 &11 &6&31 & 43 &46 &43$\pm 8$ &46-51 \\
Stishovite (eV)    &0.11 &0.52 &1.33&0.41 &0.49& 0.49&0.51$\pm 0.02$&0.51-0.54 \\
\end{tabular}
\end{ruledtabular}
\label{tab4}
\end {table*}
\begin{table*}[ht]
\caption{\label{tab:spoints} Binding energy of the 10 Smith stationary points with a hybrid functional HYB$_{\rm opt}$ (35\% of exchange), the same hybrid functional with a TS vdW correction, RPA, RPAx, SOSEX and CCSD(T) \cite{ccsdtSP} (meV). Mean absolute (percentual) error is calculated with the CCSD(T) results as reference.}
\begin{ruledtabular}
\begin{tabular}{l r r r r r r r}
  SP    & HYB$_{\rm opt}$ & HYB$_{\rm opt}$+ vdW & RPA & RPAx & SOSEX& CCSD(T)&  \\
\hline
1       	   & 216.8 & 230.0 &190.4 & 215.9 & 224.0 & 217.6\\
2              & 193.7 & 205.5 &169.7  &  194.5  &202.2& 195.6\\
3       	   & 189.9 & 201.3 &166.3 & 191.7  &199.5 &192.6 \\
4       	   & 169.8 & 184.1 &157.9  &182.1  &190.4 &187.6\\
5       	   & 157.6 & 169.3 &149.0  &171.7  &178.5 &176.6 \\
6       	   & 149.9 & 162.0 &145.3  &167.8  &173.6  &174.6 \\
7       	   & 124.6 & 138.9 &122.4 & 139.5 &  144.0 &138.6 \\
8       	   & 50.7  &  77.2 & 49.8 & 60.0   &62.8 &63.6 \\
9       	   & 131.5 & 141.1 &127.3 & 143.6  &147.4 &140.6\\
10       	   & 93.2  &  103.2    & 91.0 & 103.6  &106.3  & 100.6\\
\hline
MAE           &   11.0    & 6.9  & 21.9  & 3.1  & 4.2 & - &\\
MA\%E           &  8.1  &  5.2  &14.0      &  2.3  &2.8 &-&
\end{tabular}
\end{ruledtabular}
\label{tab5}
\end {table*}
\begin{table*}[ht]
\caption{\label{tab:struc-par-vdw} Relative energies of 10 Smith stationary points with respect to SP1 using the hybrid functional HYB$_{\rm opt}$ (35\% of exchange), the same hybrid functional with a TS vdW correction, RPA, RPAx, SOSEX and CCSD(T) \cite{ccsdtSP} (meV). Mean absolute error is calculated with the CCSD(T) results as reference.}
\begin{ruledtabular}
\begin{tabular}{l r r r r r r r}
 SP  & HYB$_{\rm opt}$ & HYB$_{\rm opt}$+ vdW & RPA & RPAx  & SOSEX & CCSD(T)  \\
\hline
1       	 & 0    & 0 & 0  & 0   & 0  & 0 \\
2            & 23.1 & 24.5 & 20.7 & 21.4&   21.8 & 22\\
3       	 & 26.9 & 28.7 & 24.1 & 24.2   & 24.4 &25\\
4       	 & 47.0   & 45.9 &32.4 &33.9&  33.6 &30\\
5       	 & 59.2 & 60.7 & 41.4 &44.2 & 45.4 &41\\
6       	 & 66.9 & 68.0 & 45.1 & 48.2   & 50.4 &43\\
7       	 & 92.2 & 91.1 & 68.0 & 76.4 & 80.0 &78\\
8       	 & 166.1& 152.8 & 140.6 & 155.9 & 161.2 &154\\
9       	 & 85.3 & 88.9 & 63.1 &72.4 & 76.6 &77\\
10       	 & 123.6& 126.8 & 99.4 & 112.4  & 117.7 &117\\
\hline
MAE           &   11.4    &  11.3 &6.8    &2.9 &2.8& - \\
\end{tabular}
\end{ruledtabular}
\label{tab6}
\end {table*}


\end{document}